\documentclass[12pt]{article}
\pdfoutput=1

\usepackage{amsmath}
\usepackage{amssymb}
\usepackage{epsfig}
\usepackage{epstopdf}
\usepackage{latexsym}
\usepackage{color}
\numberwithin{equation}{section}
\usepackage[
      colorlinks=false,
      linkcolor=darkblue,  
      urlcolor=blue,    
      filecolor=blue,     
      citecolor=red,
linktocpage=true,
      pdfstartview=FitV,
      bookmarksopen=true    
      ]{hyperref}

\begin{document}

\begin{titlepage}

\centerline
\centerline
\centerline
\bigskip
\centerline{\Huge \rm D3-branes and M5-branes wrapped} 
\bigskip
\centerline{\Huge \rm on a topological disc} 
\bigskip
\bigskip
\bigskip
\bigskip
\bigskip
\bigskip
\bigskip
\bigskip
\centerline{\rm Minwoo Suh}
\bigskip
\centerline{\it Department of Physics, Kyungpook National University, Daegu 41566, Korea}
\bigskip
\centerline{\tt minwoosuh1@gmail.com} 
\bigskip
\bigskip
\bigskip
\bigskip
\bigskip
\bigskip
\bigskip

\begin{abstract}
\noindent Employing the method applied to construct $AdS_5$ solutions from M5-branes recently by Bah, Bonetti, Minasian and Nardoni, we construct supersymmetric $AdS_3$ solutions from D3-branes and M5-branes wrapped on a disc with non-trivial holonomies at the boundary. In five-dimensional $U(1)^3$-gauged $\mathcal{N}=2$ supergravity, we find $\mathcal{N}=(2,2)$ and $\mathcal{N}=(4,4)$ supersymmetric $AdS_3$ solutions. We uplift the solutions to type IIB supergravity and obtain D3-branes wrapped on a topological disc. We also uplift the solutions to eleven-dimensional supergravity and obtain M5-branes wrapped on a product of topological disc and Riemann surface. For the $\mathcal{N}=(2,2)$ solution, holographic central charges are finite and well-defined. On the other hand, we could not find $\mathcal{N}=(4,4)$ solution with finite holographic central charge. Finally, we show that the topological disc we obtain is, in fact, identical to a special case of the multi-charge spindle solution.
\end{abstract}

\vskip 4cm

\flushleft {August, 2021}

\end{titlepage}

\tableofcontents

\bigskip
\bigskip

\section{Introduction}

A multitude of concrete and enlightening examples of the AdS/CFT correspondence, \cite{Maldacena:1997re}, have been derived via topological twist in field theories, \cite{Witten:1988ze, Bershadsky:1995vm, Bershadsky:1995qy}, and in supergravity, \cite{Maldacena:2000mw}. New field theories were constructed from their higher dimensional parents and provided the precision AdS/CFTs by holographic RG flows across dimensions.

Recently, our understanding of the AdS/CFT correspondence beyond the topological twisting has been initiated in a number of directions. In the work of \cite{Ferrero:2020laf}, in five-dimensional minimal gauged supergravity, supersymmetric $AdS_3$ solutions were constructed from D3-branes wrapped on a spindle. Although the solution was originally found in \cite{Gauntlett:2006af, Kunduri:2006uh}, the interpretation of D3-branes wrapped on a spindle was new. Spindle is topologically a two-sphere but with conical deficit angles, $2\pi\left(1-1/n_{\mp}\right)$, at the poles, where $n_{\pm}$ are two coprime positive integers. To be particular, the spindles are weighted projective space, $\mathbb{WCP}^1_{[n_-,n_+]}$. Their dual 2d $\mathcal{N}=(0,2)$ SCFTs were identified and, by employing anomaly polynomials, their central charges were matched with the gravitational results. 

In \cite{Hosseini:2021fge, Boido:2021szx}, in five-dimensional $U(1)^3$-gauged $\mathcal{N}=2$ supergravity, supersymmetric $AdS_3$ solutions from D3-branes wrapped on a multi-charge spindle were constructed which generalize the solutions of \cite{Ferrero:2020laf} in minimal gauged supergravity. The solutions were originally discovered in the study of superstar, \cite{Cvetic:1999xp, Gauntlett:2006ns}, and near-horizon of black rings, \cite{Kunduri:2007qy}. The solutions were uplifted to type IIB and eleven-dimensional supergravity and holographic central charges were matched with the field theory result from anomaly polynomials. See \cite{Hosseini:2021fge} also for further generalization to rotating and electrically charged spindle from D3-branes.

Following these works, there has been generalizations to diverse branes in M-theory. Accelerating and supersymmetric black holes were constructed from M2-branes wrapped on a spindle and their thermodynamics was studied in \cite{Ferrero:2020twa, Cassani:2021dwa}. Supersymmetric $AdS_5$ solutions from M5-branes wrapped on a spindle were studied in \cite{Ferrero:2021wvk}. 

Inspired by the line of works from \cite{Ferrero:2020laf}, a new way of preserving supersymmetry for solutions of gauged supergravity has been suggested by Bah, Bonetti, Minasian and Nardoni in \cite{Bah:2021mzw, Bah:2021hei}. In \cite{Bah:2021mzw, Bah:2021hei}, new supersymmetric $AdS_5$ solutions were constructed from M5-branes wrapping a disc with a non-trivial $U(1)$ holonomy at the boundary. Their uplift in eleven-dimensional supergravity has singularities which correspond to the M5-brane sources in the internal space. The M5-brane sources realize the irregular punctures in the dual field theory. Furthermore, the holographic dual of the solutions was proposed to be Argyres-Douglas theories from M5-branes wrapped on spheres with irregular punctures, \cite{Argyres:1995jj},  and matched central charge with the gravitational results. The method of construction was applied to find $AdS_4$ solutions from D4-D8-branes, \cite{Suh:2021aik}, and $AdS_2$ solutions from M2-branes, \cite{Suh:2021hef}. See \cite{Couzens:2021rlk} also for the study on spindle and disc solutions from M2-branes. To recapitulate, topological disc and spindle are not manifolds with constant curvature and the supersymmetry is not realized by topological twist. 

In this work, we apply the method of \cite{Bah:2021mzw, Bah:2021hei} to D3-branes and M5-branes wrapped on a disc with non-trivial holonomies at the boundary. In particular, in five-dimensional $U(1)^3$-gauged $\mathcal{N}=2$ supergravity, \cite{Cvetic:1999xp, Maldacena:2000mw}, we construct two classes of supersymmtric $AdS_3$ solutions with one $U(1)$ holonomy for $\mathcal{N}=(4,4)$ supersymmetry and with two $U(1)$ holonomies for $\mathcal{N}=(2,2)$ supersymmetry. Then, we uplift the solutions to type IIB, \cite{Schwarz:1983qr, Howe:1983sra}, and eleven-dimensional supergravity, \cite{Cremmer:1978km}. We also calculate the holographic central charges. We obtain well-defined and finite holographic central charges for the $\mathcal{N}=(2,2)$ supersymmetric $AdS_3$ solutions. However, for the $\mathcal{N}=(4,4)$ supersymmetric $AdS_3$ solutions we found, the holographic central charges diverge. Thus, holographically, unlike 2d $\mathcal{N}=(2,2)$ SCFTs, we suspect that 2d $\mathcal{N}=(4,4)$ SCFTs are not well-defined. Our solutions generalize the seminal $AdS_3$ solutions of Maldacena and N\'u\~nez, \cite{Maldacena:2000mw}, from twisted compactifications on a Riemann surface. 2d $\mathcal{N}=(4,4)$ SCFTs were shown to be not well-defined also there in \cite{Maldacena:2000mw}.

Lastly, we would like to mention that we largely follow the work of \cite{Bah:2021mzw, Bah:2021hei} and the solutions are also quite parallel in detail. We summarize the comparison of solutions in \eqref{compar} and \eqref{compar2}.

In section 2, we review five-dimensional gauged $\mathcal{N}=2$ supergravity coupled to two vector multiplets. In section 3, we construct $\mathcal{N}=(2,2)$ supersymmetric $AdS_3$ solutions. In section 4, we uplift the solutions to type IIB supergravity in order to study their geometry and calculate the holographic central charge. In section 5, we uplift the solutions to eleven-dimensional supergravity and calculate the holographic central charge. In section 6, we construct $\mathcal{N}=(4,4)$ supersymmetric $AdS_3$ solutions. However, we show that the holographic central charges diverge for the $\mathcal{N}=(4,4)$ supersymmetric solutions. In section 7, we conclude and discuss some open questions. Appendix A presents the equations of motion for five-dimensional gauged $\mathcal{N}=2$ supergravity coupled to two vector multiplets and type IIB supergravity.

\bigskip

{\bf Note added:} When we were finalizing the manuscript, the preprint, \cite{Couzens:2021tnv}, appeared on arXiv. It also studies $\mathcal{N}=(2,2)$ supersymmetric $AdS_3$ solutions from D3-branes wrapped on a topological disc. 

In appendix B, we show that the topological disc we obtain in \eqref{metmet} is, in fact, identical to the special case, \cite{Couzens:2021tnv}, of the multi-charge spindle in \cite{Hosseini:2021fge, Boido:2021szx}.

\vspace{2cm}

\section{Five-dimensional $U(1)^3$-gauged $\mathcal{N}\,=\,2$ supergravity}

We review gauged $\mathcal{N}\,=\,2$ supergravity coupled to two vector multiplets in five dimensions, \cite{Maldacena:2000mw}. The bosonic field content is the metric and graviphoton from gravity multiplet and two scalar fields and two Abelian gauge fields from two vector multiplets. The Lagrangian is given by
\begin{align}
\mathcal{L}\,&=\,R-\frac{1}{2}\partial_\mu\phi_1\partial^\mu\phi_1-\frac{1}{2}\partial_\mu\phi_2\partial^\mu\phi_2+4\left(e^{\frac{\phi_1}{\sqrt{6}}+\frac{\phi_2}{\sqrt{2}}}+e^{\frac{\phi_1}{\sqrt{6}}-\frac{\phi_2}{\sqrt{2}}}+e^{-\frac{2\phi_1}{\sqrt{6}}}\right) \notag \\
&-\frac{1}{4}\left(e^{\frac{2\phi_1}{\sqrt{6}}+\frac{2\phi_2}{\sqrt{2}}}F^1_{\mu\nu}F^{1\mu\nu}+e^{\frac{2\phi_1}{\sqrt{6}}-\frac{2\phi_2}{\sqrt{2}}}F^2_{\mu\nu}F^{2\mu\nu}+e^{-\frac{4\phi_1}{\sqrt{6}}}F^3_{\mu\nu}F^{3\mu\nu}\right)+\frac{1}{4}\epsilon^{\mu\nu\rho\sigma\lambda}F^1_{\mu\nu}F^2_{\rho\sigma}A^3_\lambda\,,
\end{align}
where $F^I_{\mu\nu}\,=\,\partial_\mu{A}^I_\nu-\partial_\nu{A}^I_\mu$. The equations of motion are presented in appendix A. The supersymmetry variations of spin 3/2- and 1/2-fields are given by
\begin{align}
\delta\psi_\mu\,=&\,D_\mu\epsilon-\frac{3i}{2}V_IA^I_\mu\epsilon+\frac{1}{2}V_IX^I\Gamma_\mu\epsilon+\frac{i}{8}X_IF^I_{\nu\rho}\left(\Gamma_\mu\,^{\nu\rho}-4\delta_\mu^\nu\Gamma^\rho\right)\epsilon\,, \notag \\
\delta\lambda_i\,=&\,-\frac{i}{2}g_{ij}\partial_\mu\phi^j\Gamma^\mu\epsilon+\frac{3i}{2}V_I\partial_iX^I\epsilon+\frac{3}{8}\partial_iX_IF^I_{\mu\nu}\Gamma^{\mu\nu}\epsilon\,,
\end{align}
where we define
\begin{equation}
g_{ij}\,=\,\frac{1}{2}\delta_{ij}\,, \qquad V_I\,=\,\frac{1}{3}\,,
\end{equation}
and $I,\,J=\,1,\,2,\,3$, $i,\,j\,=\,1,\,2$. We also employ a parametrization of scalar fields,
\begin{equation}
X^1\,=\,e^{-\frac{\phi_1}{\sqrt{6}}-\frac{\phi_2}{\sqrt{2}}}\,, \qquad X^2\,=\,e^{-\frac{\phi_1}{\sqrt{6}}+\frac{\phi_2}{\sqrt{2}}}\,, \qquad X^3\,=\,e^{\frac{2\phi_1}{\sqrt{6}}}\,,
\end{equation}
and
\begin{equation}
X_I\,=\frac{1}{3X^I}\,.
\end{equation}

\section{$\mathcal{N}\,=\,(2,2)$ supersymmetric $AdS_3$ solutions}

\subsection{Supersymmetry equations}

We consider the background,
\begin{equation}
ds^2\,=\,f(y)ds_{AdS_3}^2+g_1(y)dy^2+g_2(y)dz^2\,,
\end{equation}
with the gauge fields,
\begin{equation}
A^1\,=\,A^2\,=\,\frac{1}{2}A_z(y)dz\,, \qquad A^3\,=\,0\,,
\end{equation}
and the scalar fields,
\begin{equation}
\phi_1\,=\,\varphi(y)\,, \qquad \phi_2\,=\,0\,.
\end{equation}
The gamma matrices are given by
\begin{equation}
\Gamma^{\alpha}\,=\,\rho^{\alpha}\otimes\sigma^3\,, \qquad \Gamma^{\hat{y}}\,=\,1\otimes\sigma^1\,, \qquad \Gamma^{\hat{z}}\,=\,1\otimes\sigma^2\,,
\end{equation}
where $\alpha$ are three-dimensional flat indices and the hatted indices are flat indiced for the corresponding coordinates. $\rho^\alpha$ are three-dimensional gamma matrices with $\{\rho^\alpha,\rho^\beta\}\,=\,2\eta^{\alpha\beta}$ and $\sigma^{1,2,3}$ are the Pauli matrices. The spinor is given by
\begin{equation}
\epsilon\,=\,\vartheta\otimes\eta\,,
\end{equation}
where $\vartheta$ is a Killing spinor on $AdS_3$ and $\eta\,=\,\eta(y,z)$. The Killing spinors satisfy
\begin{equation}
\nabla_\alpha^{AdS_3}\vartheta\,=\,\frac{1}{2}s\rho_\alpha\vartheta\,,
\end{equation}
where $s\,=\,\pm1$.

The supersymmetry equations are obtained by setting the supersymmetry variations of the
fermionic fields to zero. From the supersymmetry variations, we obtain
\begin{align}
0\,=\,&-\frac{1}{2}s\Gamma^{\hat{t}}\epsilon+\frac{1}{2}f^{1/2}g_1^{-1/2}\frac{1}{2}\frac{f'}{f}\Gamma^{\hat{y}}\epsilon+\frac{1}{6}\left(2e^{-\frac{\varphi}{\sqrt{6}}}+e^{\frac{2\varphi}{\sqrt{6}}}\right)f^{1/2}\epsilon+\frac{i}{12}e^{\frac{\varphi}{\sqrt{6}}}A_z'g_1^{-1/2}g_2^{-1/2}f^{1/2}\Gamma^{\hat{y}\hat{z}}\epsilon\,, \notag \\
0\,=\,&\partial_y\epsilon+\frac{1}{6}\left(2e^{-\frac{\varphi}{\sqrt{6}}}+e^{\frac{2\varphi}{\sqrt{6}}}\right)g_1^{1/2}\Gamma^{\hat{y}}\epsilon-\frac{i}{6}e^{\frac{\varphi}{\sqrt{6}}}A_z'g_2^{-1/2}\Gamma^{\hat{z}}\epsilon\, \notag \\
0\,=\,&\partial_z\epsilon-\frac{i}{2}A_z\epsilon-\frac{1}{2}g_1^{-1/2}g_2^{-1/2}\frac{1}{2}\frac{g_2'}{g_2}\Gamma^{\hat{y}\hat{z}}\epsilon+\frac{1}{6}\left(2e^{-\frac{\varphi}{\sqrt{6}}}+e^{\frac{2\varphi}{\sqrt{6}}}\right)g_2^{1/2}\Gamma^{\hat{z}}\epsilon+\frac{i}{6}e^{\frac{\varphi}{\sqrt{6}}}g_1^{-1/2}A_z'\Gamma^{\hat{y}}\epsilon\,, \notag \\
0\,=\,&\frac{1}{2\sqrt{6}}g_1^{-1/2}\varphi'\Gamma^{\hat{y}}\epsilon+\frac{1}{3}\left(e^{-\frac{\varphi}{\sqrt{6}}}-e^{\frac{2\varphi}{\sqrt{6}}}\right)\epsilon+\frac{i}{12}e^{\frac{\varphi}{\sqrt{6}}}A_z'g_1^{-1/2}g_2^{-1/2}\Gamma^{\hat{y}\hat{z}}\epsilon\,,
\end{align}
where the first three and the last equations are from the spin-3/2 and spin-1/2 field variations, respectively. By multiplying suitable functions and gamma matrices and adding the last equation to the first three equations, we obtain
\begin{align} \label{presusy}
0\,=\,&-\frac{1}{2}s\Gamma^{\hat{t}}\epsilon+\frac{1}{2}f^{1/2}g_1^{-1/2}\frac{1}{2}\frac{f'}{f}\Gamma^{\hat{y}}\epsilon-\frac{1}{2}f^{1/2}g_1^{-1/2}\frac{1}{\sqrt{6}}\varphi'\Gamma^{\hat{y}}\epsilon+\frac{1}{2}e^{\frac{2\varphi}{\sqrt{6}}}f^{1/2}\epsilon\,, \notag \\
0\,=\,&\partial_y\epsilon-\frac{1}{2\sqrt{6}}\varphi'\epsilon+\frac{1}{2}e^{\frac{2\varphi}{\sqrt{6}}}g_1^{-1/2}\Gamma^{\hat{y}}\epsilon-\frac{i}{4}e^{-\frac{\varphi}{\sqrt{6}}}A_z'g_2^{-1/2}\Gamma^{\hat{z}}\epsilon\,, \notag \\
0\,=\,&\partial_z\epsilon-\frac{i}{2}A_z\epsilon+\frac{i}{4}e^{\frac{\varphi}{\sqrt{6}}}A_z'g_1^{-1/2}\Gamma^{\hat{y}}\epsilon+\frac{1}{2}e^{\frac{2\varphi}{\sqrt{6}}}g_2^{1/2}\Gamma^{\hat{z}}\epsilon-\frac{1}{2}g_1^{-1/2}g_2^{1/2}\frac{1}{2}\frac{g_2'}{g_2}\Gamma^{\hat{y}\hat{z}}\epsilon+\frac{1}{2}g_1^{-1/2}g_2^{1/2}\frac{1}{\sqrt{6}}\varphi'\Gamma^{\hat{y}\hat{z}}\epsilon\,, \notag \\
0\,=\,&\frac{2}{3}\left(e^{-\frac{\varphi}{\sqrt{6}}}-e^{\frac{2\varphi}{\sqrt{6}}}\right)\epsilon+\frac{1}{\sqrt{6}}g_1^{-1/2}\varphi'\Gamma^{\hat{y}}\epsilon+\frac{i}{6}e^{\frac{\varphi}{\sqrt{6}}}A_z'g_1^{-1/2}g_2^{-1/2}\Gamma^{\hat{y}\hat{z}}\epsilon\,.
\end{align}
The spinor is supposed to have a charge under the $U(1)_z$ isometry,
\begin{equation}
\eta(y,z)\,=\,e^{inz}\widehat{\eta}(y)\,,
\end{equation}
where $n$ is a constant. It shows up in the supersymmetry equations in the form of $\left(-i\partial_z+\frac{1}{2}A_z\right)\eta\,=\,\left(n+\frac{1}{2}A_z\right)\eta$ which is invariant under
\begin{equation}
A^1\,\mapsto\,A^1-2\alpha_0dz\,, \qquad A^2\,\mapsto\,A^2-2\alpha_0dz\,, \qquad \eta\,\mapsto\,e^{i\alpha_0z}\eta\,,
\end{equation}
where $\alpha_0$ is a constant. We also define
\begin{equation} \label{AAhat}
\frac{1}{2}\widehat{A}_z\,=\,n+\frac{1}{2}A_z\,.
\end{equation}
We solve the equation of motion for the gauge fields and obtain
\begin{equation} \label{solgaugep}
A_z'\,=\,be^{-\frac{2\varphi}{\sqrt{6}}}g_1^{1/2}g_2^{1/2}f^{-3/2}\,,
\end{equation}
where $b$ is a constant. Employing the expressions we discussed in \eqref{presusy} beside the second equation, we finally obtain the supersymmetry equations,
\begin{align}
0\,=\,&-sf^{-1/2}\eta+g_1^{-1/2}\left[\frac{1}{2}\frac{f'}{f}-\frac{1}{\sqrt{6}}\varphi'\right]\left(i\sigma^2\eta\right)+e^{\frac{2\varphi}{\sqrt{6}}}\left(\sigma^3\eta\right)\,, \notag \\
0\,=\,&g_2^{-1/2}\widehat{A}_z\left(\sigma^1\eta\right)-\frac{1}{2}f^{-3/2}e^{-\frac{\varphi}{\sqrt{6}}}\,b\,\eta-g_1^{-1/2}\left[\frac{1}{2}\frac{g_2'}{g_2}-\frac{1}{\sqrt{6}}\varphi'\right]\left(i\sigma^2\eta\right)-e^{\frac{2\varphi}{\sqrt{6}}}\left(\sigma^3\eta\right)\,, \notag \\
0\,=\,&\frac{1}{3}\left(e^{-\frac{\varphi}{\sqrt{6}}}-e^{\frac{2\varphi}{\sqrt{6}}}\right)\left(\sigma^3\eta\right)+\frac{1}{2\sqrt{6}}g_1^{-1/2}\varphi'\left(i\sigma^2\eta\right)-\frac{1}{12}f^{-3/2}e^{-\frac{\varphi}{\sqrt{6}}}\,b\,\eta\,.
\end{align}

The supersymmetry equations are in the form of $M^{(i)}\eta\,=\,0$, $i\,=\,1,\,2,\,3$, where $M^{(i)}$ are three $2\times{2}$ matrices, as we follow \cite{Bah:2021mzw, Bah:2021hei},
\begin{equation}
M^{(i)}\,=\,X_0^{(i)}\mathbb{I}_2+X_1^{(i)}\sigma^1+X_2^{(i)}\left(i\sigma^2\right)+X_3^{(i)}\sigma^3\,.
\end{equation}
We rearrange the matrices to introduce $2\times{2}$ matrices,
\begin{equation}
\mathcal{A}^{ij}\,=\,\text{det}\left(v^{(i)}|w^{(j)}\right)\,, \qquad \mathcal{B}^{ij}\,=\,\text{det}\left(v^{(i)}|v^{(j)}\right)\,, \qquad \mathcal{C}^{ij}\,=\,\text{det}\left(w^{(i)}|w^{(j)}\right)\,,
\end{equation}
from the column vectors of
\begin{equation}
v^{(i)}\,=\,\left(
\begin{array}{l}
 X_1^{(i)}+X_2^{(i)} \\
 -X_0^{(i)}-X_3^{(i)}
\end{array}
\right)\,, \qquad
w^{(i)}\,=\,\left(
\begin{array}{l}
 X_0^{(i)}-X_3^{(i)} \\
 -X_1^{(i)}+X_2^{(i)}
\end{array}
\right)\,.
\end{equation}

From the vanishing of $\mathcal{A}^{ij}$, $\mathcal{B}^{ij}$ and $\mathcal{C}^{ij}$, necessary conditions for non-trivial solutions are obtained. From $\mathcal{A}^{ii}\,=\,0$, we find
\begin{align} \label{Adiag}
0\,=\,&\frac{1}{f}+\frac{1}{g_1}\left(\frac{1}{2}\frac{f'}{f}-\frac{1}{\sqrt{6}}\varphi'\right)^2-e^{\frac{4\varphi}{\sqrt{6}}}\,, \notag \\
0\,=\,&\frac{b^2e^{-\frac{2\varphi}{\sqrt{6}}}}{4f^3}+\frac{1}{g_1}\left(\frac{1}{2}\frac{g_2'}{g_2}-\frac{1}{\sqrt{6}}\varphi'\right)^2-e^{\frac{4\varphi}{\sqrt{6}}}-\frac{\widehat{A}_z^2}{g_2}\,, \notag \\
0\,=\,&\frac{(\varphi')^2}{24g_1}+\frac{b^2e^{-\frac{2\varphi}{\sqrt{6}}}}{144f^3}-\frac{1}{9}\left(e^{-\frac{\varphi}{\sqrt{6}}}-e^{\frac{2\varphi}{\sqrt{6}}}\right)^2\,.
\end{align}
From $\mathcal{A}^{ij}+\mathcal{A}^{ji}\,=\,0$, we find
\begin{align}
0\,=\,&\frac{2}{g_1}\left(\frac{1}{2}\frac{f'}{f}-\frac{1}{\sqrt{6}}\varphi'\right)\left(\frac{1}{2}\frac{g_2'}{g_2}-\frac{1}{\sqrt{6}}\varphi'\right)-\frac{s\,b\,e^{-\frac{\varphi}{\sqrt{6}}}}{f^2}-2e^{\frac{4\varphi}{\sqrt{6}}}\,, \notag \\
0\,=\,&-\frac{s\,b\,e^{-\frac{\varphi}{\sqrt{6}}}}{6f^2}-\frac{1}{\sqrt{6}g_1}\varphi'\left(\frac{1}{2}\frac{f'}{f}-\frac{1}{\sqrt{6}}\varphi'\right)+\frac{2}{3}e^{\frac{2\varphi}{\sqrt{6}}}\left(e^{-\frac{\varphi}{\sqrt{6}}}-e^{\frac{2\varphi}{\sqrt{6}}}\right)\,, \notag \\
0\,=\,&\frac{b^2\,e^{-\frac{2\varphi}{\sqrt{6}}}}{12f^3}-\frac{1}{\sqrt{6}g_1}\varphi'\left(\frac{1}{2}\frac{g_2'}{g_2}-\frac{1}{\sqrt{6}}\varphi'\right)+\frac{2}{3}e^{\frac{2\varphi}{\sqrt{6}}}\left(e^{-\frac{\varphi}{\sqrt{6}}}-e^{\frac{2\varphi}{\sqrt{6}}}\right)\,.
\end{align}
From $\mathcal{A}^{ij}-\mathcal{A}^{ji}\,=\,0$, we find
\begin{align} \label{Aminus}
0\,=\,&\frac{b\,e^{\frac{\varphi}{\sqrt{6}}}}{f^{3/2}}+\frac{2s\,e^{\frac{2\varphi}{\sqrt{6}}}}{\sqrt{f}}+\frac{2\hat{A}_z}{\sqrt{g_1}\sqrt{g_2}}\left(\frac{1}{2}\frac{f'}{f}-\frac{1}{\sqrt{6}}\varphi'\right)\,, \notag \\
0\,=\,&-\frac{b\,e^{\frac{\varphi}{\sqrt{6}}}}{6f^{3/2}}+\frac{2s}{3\sqrt{f}}\left(e^{-\frac{\varphi}{\sqrt{6}}}-e^{\frac{2\varphi}{\sqrt{6}}}\right)\,, \notag \\
0\,=\,&\frac{b\,e^{\frac{\varphi}{\sqrt{6}}}}{6f^{3/2}}+\frac{b\,e^{-\frac{\varphi}{\sqrt{6}}}}{3f^{3/2}}\left(e^{-\frac{\varphi}{\sqrt{6}}}-e^{\frac{2\varphi}{\sqrt{6}}}\right)+\frac{\varphi'\widehat{A}_z}{\sqrt{6}\sqrt{g_1}\sqrt{g_2}}\,.
\end{align}
From $\mathcal{B}^{ij}+\mathcal{C}^{ij}\,=\,0$, we find 
\begin{align}
0\,=\,&-\frac{b\,e^{-\frac{\varphi}{\sqrt{6}}}}{f^{3/2}\sqrt{g_1}}\left(\frac{1}{2}\frac{f'}{f}-\frac{1}{\sqrt{6}}\varphi'\right)-\frac{2s}{\sqrt{f}\sqrt{g_1}}\left(\frac{1}{2}\frac{g_2'}{g_2}-\frac{1}{\sqrt{6}}\varphi'\right)-\frac{2e^{\frac{2\varphi}{\sqrt{6}}}\widehat{A}_z}{\sqrt{g_2}}\,, \notag \\
0\,=\,&\frac{b\,e^{-\frac{\varphi}{\sqrt{6}}}}{6f^{3/2}\sqrt{g_1}}\left(\frac{1}{2}\frac{f'}{f}-\frac{1}{\sqrt{6}}\varphi'\right)-\frac{s\,\varphi'}{\sqrt{6}\sqrt{f}\sqrt{g_1}}\,, \notag \\
0\,=\,&-\frac{b\,e^{-\frac{\varphi}{\sqrt{6}}}}{6f^{3/2}\sqrt{g_1}}\left(\frac{1}{2}\frac{g_2'}{g_2}-\frac{1}{\sqrt{6}}\varphi'\right)-\frac{b\,e^{-\frac{\varphi}{\sqrt{6}}}\varphi'}{2\sqrt{6}f^{3/2}\sqrt{g_1}}-\frac{2\widehat{A}_z}{3\sqrt{g_2}}\left(e^{-\frac{\varphi}{\sqrt{6}}}-e^{\frac{2\varphi}{\sqrt{6}}}\right)\,.
\end{align}
From $\mathcal{B}^{ij}-\mathcal{C}^{ij}\,=\,0$, we find 
\begin{align} \label{Abc}
0\,=\,&\frac{2e^{\frac{2\varphi}{\sqrt{6}}}}{\sqrt{g_1}}\left(\frac{1}{2}\frac{g_2'}{g_2}-\frac{1}{\sqrt{6}}\varphi'\right)-\frac{2e^{\frac{2\varphi}{\sqrt{6}}}}{\sqrt{g_1}}\left(\frac{1}{2}\frac{f'}{f}-\frac{1}{\sqrt{6}}\varphi'\right)+\frac{2s\widehat{A}_z}{\sqrt{f}\sqrt{g_2}}\,, \notag \\
0\,=\,&-\frac{2}{3\sqrt{g_1}}\left(\frac{1}{2}\frac{f'}{f}-\frac{1}{\sqrt{6}}\varphi'\right)\left(e^{-\frac{\varphi}{\sqrt{6}}}-e^{\frac{2\varphi}{\sqrt{6}}}\right)+\frac{e^{\frac{2\varphi}{\sqrt{6}}}\varphi'}{\sqrt{6}\sqrt{g_1}}\,, \notag \\
0\,=\,&\frac{2}{3\sqrt{g_1}}\left(\frac{1}{2}\frac{g_2'}{g_2}-\frac{1}{\sqrt{6}}\varphi'\right)\left(e^{-\frac{\varphi}{\sqrt{6}}}-e^{\frac{2\varphi}{\sqrt{6}}}\right)-\frac{e^{\frac{2\varphi}{\sqrt{6}}}\varphi'}{\sqrt{6}\sqrt{g_1}}+\frac{b\,e^{-\frac{\varphi}{\sqrt{6}}}\widehat{A}_z}{6f^{3/2}\sqrt{g_2}}\,.
\end{align}

\subsection{Supersymmetric solutions}

From the second equation of \eqref{Aminus}, we obtain
\begin{equation} \label{fsol}
f\,=\,\frac{b}{4se^{-\frac{\varphi}{\sqrt{6}}}\left(e^{-\frac{\varphi}{\sqrt{6}}}-e^{\frac{2\varphi}{\sqrt{6}}}\right)}\,.
\end{equation}
Then, from the third equation of \eqref{Adiag} with \eqref{fsol}, we obtain
\begin{equation}
g_1\,=\,\frac{3b\left(\varphi'\right)^2}{8\left(e^{-\frac{\varphi}{\sqrt{6}}}-e^{\frac{2\varphi}{\sqrt{6}}}\right)^2\left(b-4se^{-\frac{5\varphi}{\sqrt{6}}}\left(e^{-\frac{\varphi}{\sqrt{6}}}-e^{\frac{2\varphi}{\sqrt{6}}}\right)\right)}\,.
\end{equation}
From the third equation of \eqref{Aminus}, we find an expression for $\sqrt{g_1}\sqrt{g_2}$,
\begin{equation} \label{g1g21}
\sqrt{g_1}\sqrt{g_2}\,=\,\frac{6^{1/2}b^{1/2}e^{\frac{5\varphi}{\sqrt{6}}}\widehat{A}_z\varphi'}{4^{3/2}\left(2-e^{\frac{3\varphi}{\sqrt{6}}}\right)\left(1-e^{\frac{3\varphi}{\sqrt{6}}}\right)^{3/2}}\,.
\end{equation}
Also from \eqref{solgaugep}, we find another expression for $\sqrt{g_1}\sqrt{g_2}$,
\begin{equation} \label{g1g22}
\sqrt{g_1}\sqrt{g_2}\,=\,\frac{b^{1/2}e^{\frac{5\varphi}{\sqrt{6}}}A_z'}{8\left(1-e^{\frac{3\varphi}{\sqrt{6}}}\right)^{3/2}}\,.
\end{equation}
Equating \eqref{g1g21} and \eqref{g1g22}, we find an ordinary differential equation for $\widehat{A}_z$ and it gives
\begin{equation}
\widehat{A}_z\,=\,\mathcal{C}e^{-\frac{2\varphi}{\sqrt{6}}}\left(2e^{-\frac{\varphi}{\sqrt{6}}}-e^{\frac{2\varphi}{\sqrt{6}}}\right)\,,
\end{equation}
where $\mathcal{C}$ is a constant. From \eqref{AAhat}, we find
\begin{equation}
A_z\,=\,\mathcal{C}e^{-\frac{2\varphi}{\sqrt{6}}}\left(2e^{-\frac{\varphi}{\sqrt{6}}}-e^{\frac{2\varphi}{\sqrt{6}}}\right)+n\,.
\end{equation}
Then, from \eqref{g1g21} or \eqref{g1g22}, we obtain 
\begin{equation}
g_2\,=\,\frac{\mathcal{C}^2\left(b-4se^{-\frac{5\varphi}{\sqrt{6}}}\left(e^{-\frac{\varphi}{\sqrt{6}}}-e^{\frac{2\varphi}{\sqrt{6}}}\right)\right)}{4se^{-\frac{\varphi}{\sqrt{6}}}\left(e^{-\frac{\varphi}{\sqrt{6}}}-e^{\frac{2\varphi}{\sqrt{6}}}\right)}\,.
\end{equation}
Therefore, we have determined all functions in terms of the scalar field, $\varphi(y)$, and its derivative. The solution satisfies all the supersymmetry equations in \eqref{Adiag} to \eqref{Abc} and the equations motion which we present in appendix A. We can determine the scalar field by fixing the ambiguity in reparametrization of $r$ due to the covariance of the supersymmetry equations,
\begin{equation}
\varphi(y)\,=\,\frac{2\sqrt{6}}{3}\log{y}\,,
\end{equation}
where $y\,>\,0$.

Finally, let us summarize the solution. The metric is given by
\begin{equation} \label{metmet}
ds^2\,=\,\frac{by^{4/3}}{4s\left(1-y^2\right)}\left[ds_{AdS_3}^2+\frac{4}{sy^2\left(1-y^2\right)h(y)}dy^2+\frac{\mathcal{C}^2h(y)}{b}dz^2\right]\,,
\end{equation}
where we define
\begin{equation}
h(y)\,=\,b-4sy^{-4}\left(1-y^2\right)\,.
\end{equation}
The gauge field is given by
\begin{equation}
\widehat{A}_z\,=\,\mathcal{C}\left(2y^{-2}-1\right)\,.
\end{equation}
The metric can also be written as
\begin{equation}
ds^2\,=\,\frac{by^{4/3}}{4s\left(1-y^2\right)}ds_{AdS_3}^2+\frac{by^{-2/3}}{h(y)\left(1-y^2\right)^2}dy^2+\frac{\mathcal{C}^2y^{4/3}h(y)}{4s\left(1-y^2\right)}dz^2\,.
\end{equation}

Now we consider the range of $y$ for regular solutions, $i.e.$, the metric functions are positive definite and the scalar fields are real.  We find regular solutions when we have
\begin{equation} \label{regrange}
s\,=\,1\,, \qquad -1<b<0\,, \qquad 1<y_1<y<\infty\,, \qquad y_1\,=\,\sqrt{-\frac{2\left(1+\sqrt{1+b}\right)}{b}}\,.
\end{equation}
where $y_1$ is determined from $h(y_1)\,=\,0$. We plot a representative solution with $b\,=\,-0.5$ and $\mathcal{C}\,=\,1$ in Figure 1. The metric on the space spanned by $\Sigma(y,z)$ in \eqref{metmet} has a topology of disc with the origin at $y\,=\,y_1$ and the boundary at $y\,=\,0$.

\begin{figure}[t]
\begin{center}
\includegraphics[width=2.0in]{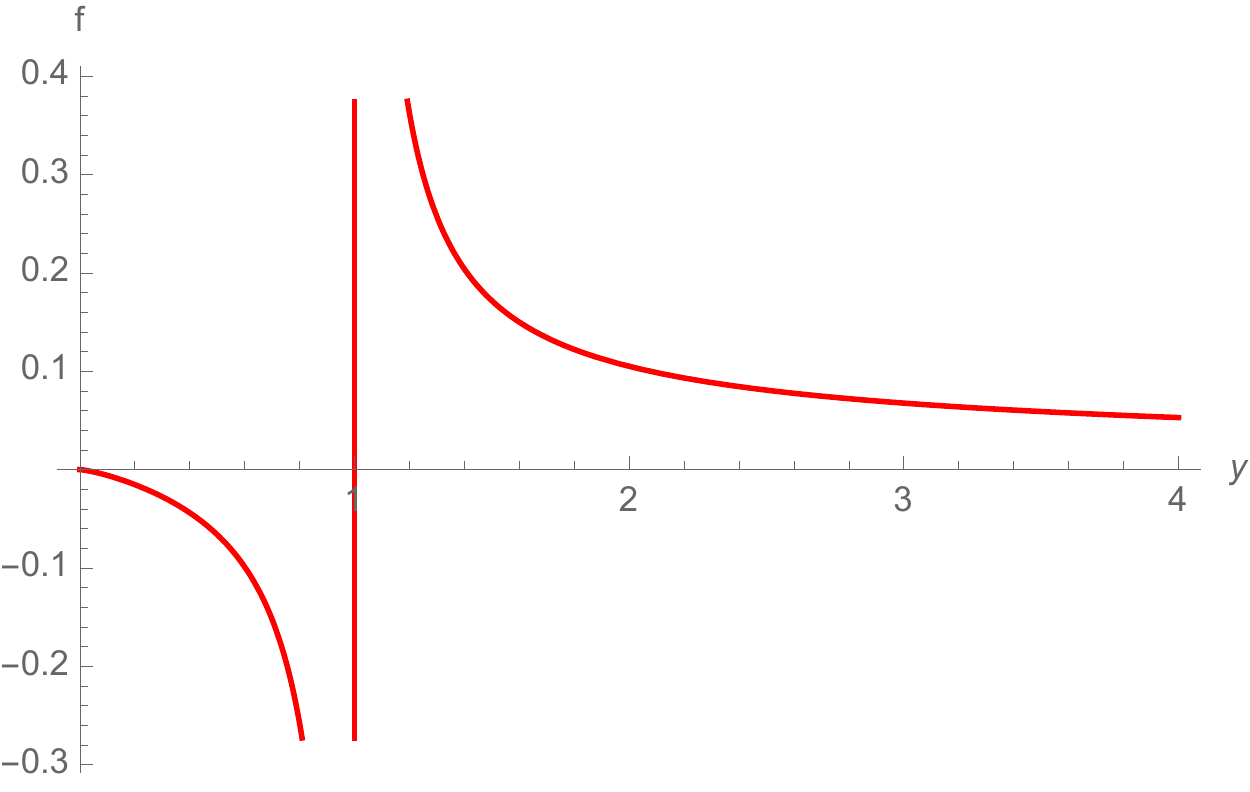} \qquad \includegraphics[width=2.0in]{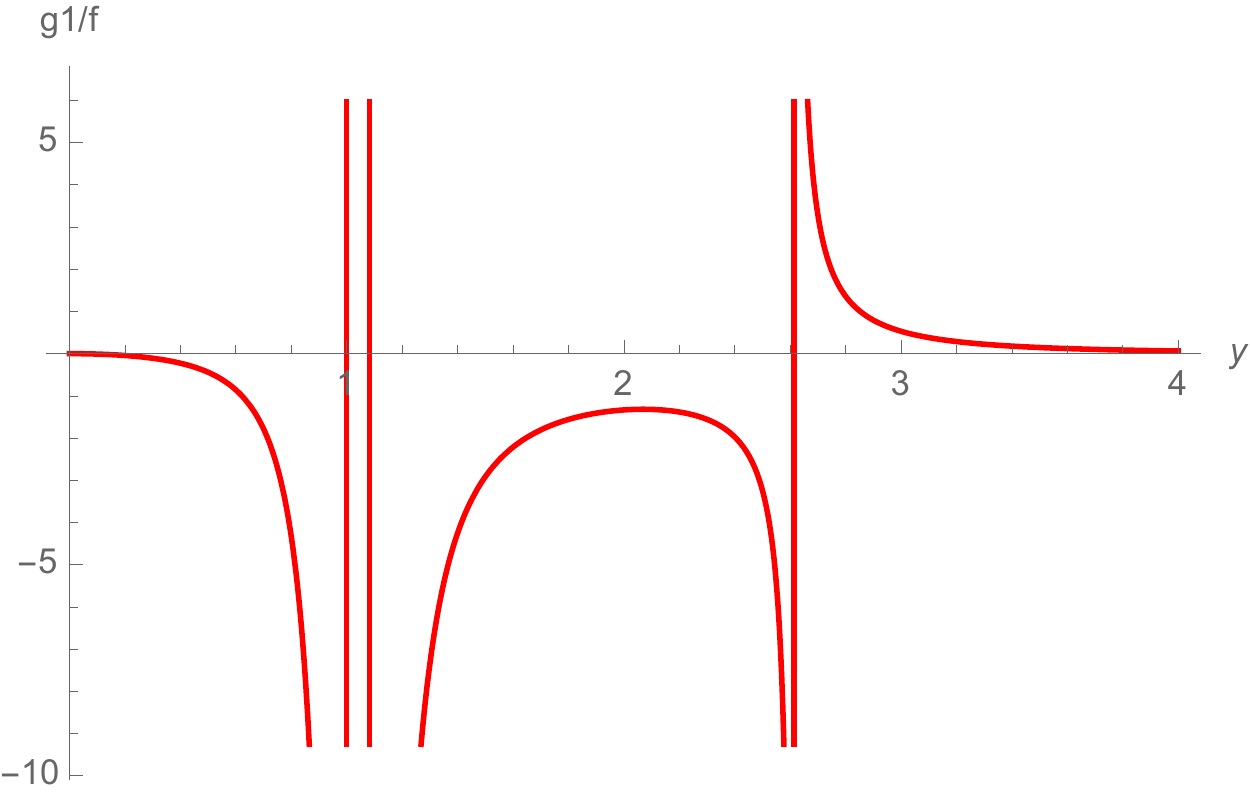} \qquad \includegraphics[width=2.0in]{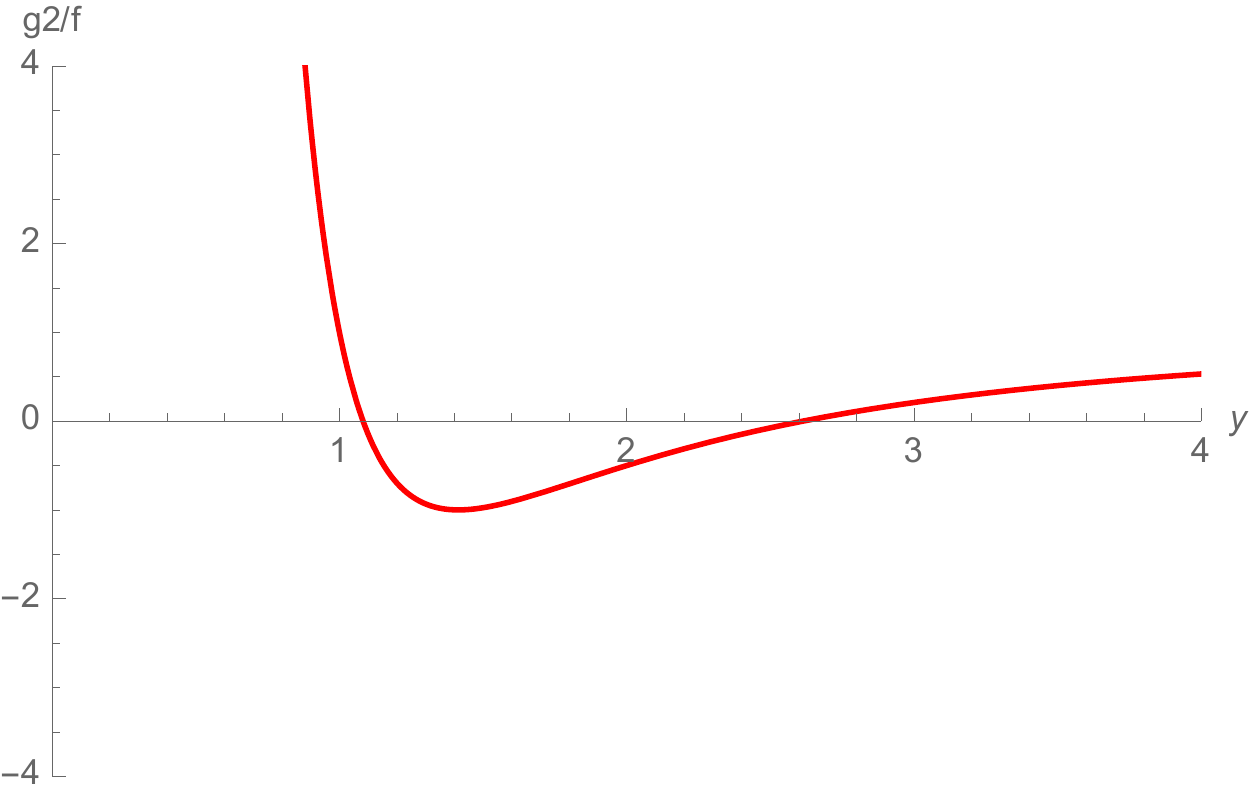}
\caption{{\it A representative solution with $b=-0.5$ and $\mathcal{C}\,=\,1$. The solution is regular in the range of $y_1=2.6131\,<\,y\,<\,\infty$.}}
\end{center}
\end{figure}

Near $y\,\rightarrow\,\infty$ the $AdS_3$ warp factor vanishes and it is a curvature singularity of the metric,
\begin{equation} \label{wsingu}
ds^2\,\approx\,\frac{b}{4sy^{2/3}}\left[ds_{AdS_3}^2+\frac{4}{-sby^4}dy^2+\mathcal{C}^2dz^2\right]\,.
\end{equation}
This singularity is resolved when the solution is uplifted to type IIB or eleven-dimensional supergravity.

Approaching $y\,=\,y_1$, the metric becomes to be 
\begin{equation}
ds^2\,=\,\frac{by_1^{4/3}}{4s\left(1-y_1^2\right)}\left[ds_{AdS_3}^2+\frac{16\Big[d\rho^2+\mathcal{C}^2\left(1+b\right)\rho^2dz^2\Big]}{-h'(y_1)y_1^2\left(1-y_1^2\right)}\right]\,,
\end{equation}
where we introduced a new parametrization of coordinate, $\rho^2\,=\,y_1-y$. Then, the $\rho$-$z$ surface is locally an $\mathbb{R}^2/\mathbb{Z}_l$ orbifold if we set
\begin{equation} \label{clb}
\mathcal{C}\,=\,\frac{1}{l\sqrt{1+b}}\,,
\end{equation}
where $l\,=\,1,\,2,\,3,\ldots\,\,$.

Employing the Gauss-Bonnet theorem, we calculate the Euler characteristic of $\Sigma$, the $y$-$z$ surface, from \eqref{metmet}. The boundary at $y\,=\,1$ is a geodesic and thus has vanishing geodesic curvature. The only contribution to the Euler characteristic is  
\begin{equation}
\chi\left(\Sigma\right)\,=\,\frac{1}{4\pi}\int_\Sigma{R}_\Sigma\text{vol}_\Sigma\,=\,\frac{2\pi}{4\pi}\frac{4\mathcal{C}\left(y_1^2-2\right)\sqrt{b\left(1-y_1^2\right)}}{by^4}\,=\,\mathcal{C}\sqrt{1+b}\,=\,\frac{1}{l}\,,
\end{equation}
where $0\,<\,z\,<2\pi$. This result is natural for a disc in an $\mathbb{R}^2/\mathbb{Z}_l$ orbifold at $y\,=\,1$.

\section{D3-branes wrapped on a topological disc}

\subsection{Uplift to type IIB supergravity}

We uplift the solution to type IIB supergravity, \cite{Schwarz:1983qr, Howe:1983sra}. The only non-trivial fields in type IIB supergravity are the metric and the RR five-form flux. The uplift formula, \cite{Cvetic:1999xp}, is given for the metric,
\begin{equation}
ds_{10}^2\,=\,\sqrt{\Delta}ds_5^2+\frac{1}{\sqrt{\Delta}}\sum_{I=1}^3\frac{1}{X^I}\left(d\mu_I^2+\mu_I^2\left(d\phi_I+A^I\right)\right)\,,
\end{equation}
and for the self-dual five-form flux, $F_{(5)}\,=\,G_{(5)}+*G_{(5)}$,{\footnote{A sign error in the second term of the formula in \cite{Cvetic:1999xp} is corrected in \cite{Gauntlett:2006ns} and \cite{Benini:2013cda}.}}
\begin{equation}
G_{(5)}\,=\,\sum_{I=1}^3\left[2X^I\left(X^I\mu_I^2-\Delta\right)\text{vol}_5+\frac{1}{2X^I}d\left(\mu_I^2\right)\wedge*_5dX^I+\frac{1}{2\left(X^I\right)^2}d\left(\mu_I^2\right)\wedge\left(d\phi_I+A^I\right)\wedge*_5F^I\right]\,,
\end{equation}
where we define
\begin{equation}
\Delta\,=\,\sum_{I=1}^3X^I\mu_I^2\,,
\end{equation}
and $\text{vol}_5$ and $*_5$ denote the volume form and the Hodge dual with respect to the five-dimensional metric, $ds_5^2$, respectively. The five-dimensional fields are $ds_5^2$, $F^I\,=\,dA^I$, and $X^I$. We introduce a parametrization in terms of angles on a two-sphere,
\begin{equation}
\mu_1\,=\,\cos\xi\sin\psi\,, \qquad \mu_2\,=\,\cos\xi\cos\psi\,, \qquad \mu_3\,=\,\sin\xi\,,
\end{equation}
and the ranges of the internal coordinates are{\footnote{In numerous literature, the ranges of the internal coordinates are incorrectly recorded.}}
\begin{equation}
0\le\xi\,,\psi\le\frac{\pi}{2}\,, \qquad 0\le\phi_1\,,\phi_2\,,\phi_3\le2\pi\,.
\end{equation}

By employing the uplift formula, we obtain the uplifted metric,
\begin{equation} \label{upmet}
ds_{10}^2\,=\,\sqrt{\Delta}\left[ds_5^2+e^{-\frac{\varphi}{\sqrt{6}}}d\xi^2\right] +\frac{1}{\sqrt{\Delta}}\left[e^{\frac{\varphi}{\sqrt{6}}}\cos^2\xi\left(d\psi^2+\sin^2\psi{D}\phi_1^2+\cos^2\psi{D}\phi_2^2\right)+e^{-\frac{2\varphi}{\sqrt{6}}}\sin^2\xi{D}\phi_3^2\right]\,,
\end{equation}
where we have
\begin{equation}
\Delta\,=\,e^{-\frac{\varphi}{\sqrt{6}}}\cos^2\xi+e^{\frac{2\varphi}{\sqrt{6}}}\sin^2\xi\,,
\end{equation}
and
\begin{equation}
D\phi_I\,=\,d\phi_I+A^I\,.
\end{equation}
In particular, for our solutions, the metric can also be written by
\begin{align} \label{upmet1}
ds_{10}^2\,=&\,\frac{by^{4/3}\Delta^{1/2}}{4\left(1-y^2\right)}\left[ds_{AdS_3}^2+\frac{4}{y^2\left(1-y^2\right)h}dy^2+\frac{\mathcal{C}^2h}{b}dz^2+\frac{4\left(1-y^2\right)}{by^2}d\xi^2\right. \notag \\
+&\left.\frac{4\left(1-y^2\right)}{by^{2/3}\Delta}\cos^2\xi\left(d\psi^2+\sin^2\psi{D}\phi_1^2+\cos^2\psi{D}\phi_2^2\right)+\frac{4\left(1-y^2\right)}{by^{8/3}\Delta}\sin^2\xi{d}\phi_3^2\right]\,,
\end{align}
with
\begin{equation}
\Delta\,=\,y^{-2/3}\cos^2\xi+y^{4/3}\sin^2\xi\,.
\end{equation}

We find the self-dual five-form flux, $F_{(5)}\,=\,G_{(5)}+*G_{(5)}$, to be
\begin{align}
G_{(5)}\,=\,&-\frac{b^2\mathcal{C}y^{7/3}}{8\left(1-y^2\right)^3}\left(y^{2/3}+y^{-2/3}\Delta\right)dy\wedge{d}z\wedge\text{vol}_{AdS_3} \notag \\
&-\frac{b\,\mathcal{C}y^2h}{8\left(1-y^2\right)}\cos\xi\sin\xi\,d\xi\wedge{d}z\wedge\text{vol}_{AdS_3} \notag \\
&+\frac{b}{2}\cos\xi\sin\psi\Big(\sin\xi\sin\psi{d}\xi-\cos\xi\cos\psi{d}\psi\Big)\wedge{D}\phi_1\wedge\text{vol}_{AdS_3} \notag \\
&+\frac{b}{2}\cos\xi\cos\psi\Big(\sin\xi\cos\psi{d}\xi+\cos\xi\sin\psi{d}\psi\Big)\wedge{D}\phi_2\wedge\text{vol}_{AdS_3}\,,
\end{align}
and
\begin{align}
*G_{(5)}\,=\,&-\frac{2}{\Delta^2}\left(y^{2/3}+y^{-2/3}\Delta\right)\cos^3\xi\sin\xi\,d\xi\wedge\text{vol}_{S^3}\wedge{d}\phi_3 \notag \\
&-\frac{2}{y^{1/3}\Delta^2}\cos^4\xi\sin^2\xi\,dy\wedge\text{vol}_{S^3}\wedge{d}\phi_3 \notag \\
&-\frac{2\mathcal{C}}{y^3}\cos\xi\sin\xi\,d\xi\wedge\left(\sin^2\psi{d}\phi_1+\cos^2\psi{d}\phi_2\right)\wedge{d}\phi_3\wedge{d}y\wedge{d}z \notag \\
&-\frac{2\mathcal{C}}{y^{5/3}\Delta}\cos^2\xi\sin^2\xi\cos\psi\sin\psi\,d\psi\wedge\left(d\phi_1-d\phi_2\right)\wedge{d}\phi_3\wedge{d}y\wedge{d}z\,,
\end{align}
where we introduce the volume form of the gauged three-sphere,
\begin{equation}
\text{vol}_{S^3}\,=\,\cos\psi\sin\psi\,d\psi\wedge{D}\phi_1\wedge{D}\phi_2\,.
\end{equation}
We note that the first two terms of $*G_{(5)}$ can also be written by
\begin{equation}
*G_{(5)}\,=\,-\text{vol}_{S^3}\wedge{d}\left[\frac{\cos^4\xi}{y^{2/3}\Delta}d\phi_3\right]+\ldots\,.
\end{equation}

\subsection{Uplifted metric}

The seven-dimensional internal space of the uplifted metric is an $S_z^1\,\times\,S_{\phi_3}^1\,\times\,S^3$ fibration over the 2d base space, $B_2$, of $(y,\xi)$. The three-sphere, $S^3$, is spanned by $(\psi,\phi_1,\phi_2)$. The 2d base space is a rectangle of $(y,\xi)$ over $[y_1,\infty)\,\times\left[0,\frac{\pi}{2}\right]$. See Figure 2. We explain the geometry of the internal space by three regions of the 2d base space, $B_2$.

\begin{itemize}
\item Region I: The side of $\mathsf{P}_1\mathsf{P}_2$.
\item Region II: The sides of $\mathsf{P}_2\mathsf{P}_3$ and $\mathsf{P}_3\mathsf{P}_4$.
\item Region III: The side of $\mathsf{P}_1\mathsf{P}_4$.
\end{itemize}

\begin{figure}[t]
\begin{center}
\includegraphics[width=4.5in]{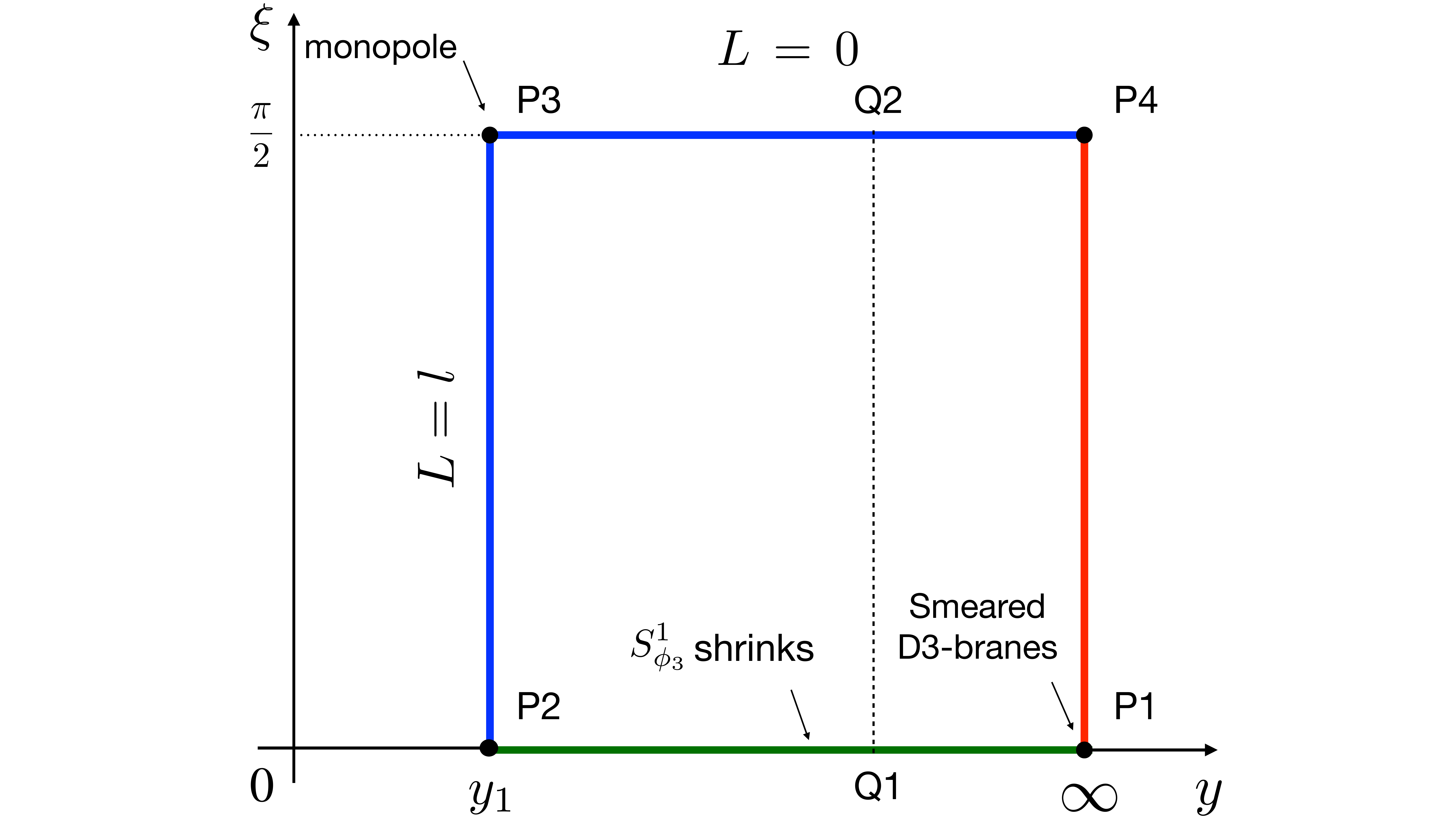}
\caption{{\it The two-dimensional base space, $B_2$, spanned by $y$ and $\xi$.}}
\end{center}
\end{figure}

\noindent {\bf Region I:} On the side of $\xi\,=\,0$, the circle, $S_{\phi_3}^1$, shrinks and the internal space caps off.

\noindent {\bf Region II: Monopole} We break $D\phi_1$ and $D\phi_2$ and complete the square of $dz$, \cite{Bah:2021hei, Couzens:2021rlk}, to obtain the metric of
\begin{align}
ds_{10}^2\,=\,\frac{by^{4/3}\Delta^{1/2}}{4\left(1-y^2\right)}&\left[ds_{AdS_3}^2+\frac{4dy^2}{y^2\left(1-y^2\right)h}+\frac{4\left(1-y^2\right)d\xi^2}{by^2}+\frac{4\left(1-y^2\right)}{by^{2/3}\Delta}\cos^2\xi{d}\psi^2\right. \notag \\
&+\frac{4\left(1-y^2\right)}{by^{8/3}\Delta}\sin^2\xi{d}\phi_3^2+R_z^2\left(dz-\frac{L}{\cos^2\xi}\left(\mu_1^2d\phi_1+\mu_2^2d\phi_2\right)\right)^2 \notag \\
&\left.+R_{\phi_1}^2\mu_1^2\left(d\phi_1-L_1\mu_2^2d\phi_2\right)^2+R_{\phi_2}^2\mu_2^2d\phi_2^2\right]\,.
\end{align}
The metric functions are defined to be
\begin{align}
R_z^2\,=&\,\frac{\mathcal{C}^2\left(\Delta\,hy^{14/3}-\left(y^2-1\right)\left(y^2-2\right)^2\left(\mu_1^2+\mu_2^2\right)\right)}{b\Delta\,y^{14/3}}\,, \notag \\
R_{\phi_1}^2\,=&\,\frac{-4\left(y^2-1\right)\left(\Delta\,hy^{14/3}-\left(y^2-1\right)\left(y^2-2\right)^2\mu_2^2\right)}{b\Delta\,y^{2/3}\left(\Delta\,hy^{14/3}-\left(y^2-1\right)\left(y^2-2\right)^2\left(\mu_1^2+\mu_2^2\right)\right)}\,, \notag \\
R_{\phi_2}^2\,=&\,\frac{-4y^4\left(y^2-1\right)h}{b\left(\Delta\,hy^{14/3}-\left(y^2-1\right)\left(y^2-2\right)^2\mu_2^2\right)}\,,
\end{align}
with
\begin{align}
L\,=&\,\frac{-2y^2\left(y^2-1\right)\left(y^2-2\right)\cos^2\xi}{\mathcal{C}\left(\Delta\,hy^{14/3}-\left(y^2-1\right)\left(y^2-2\right)^2\left(\mu_1^2+\mu_2^2\right)\right)}\,, \notag \\
L_1\,=&\,\frac{-\left(y^2-1\right)\left(y^2-2\right)^2}{\Delta\,hy^{14/3}-\left(y^2-1\right)\left(y^2-2\right)^2\mu_2^2}\,.
\end{align}

The function, $L(y,\xi)$, is piecewise constant along the sides of $y\,=\,y_1$ and $\xi\,=\,\frac{\pi}{2}$ of the 2d base, $B_2$,
\begin{equation}
L\left(y,\frac{\pi}{2}\right)\,=\,0\,, \qquad L\left(y_1,\xi\right)\,=\,\frac{1}{\mathcal{C}\sqrt{1+b}}\,.
\end{equation}
The jump in $L$ at the corner, $(y,\xi)\,=\,\left(y_1,\frac{\pi}{2}\right)$, indicates the existence of a monopole source for the $Dz$ fibration.

We perform a coordinate transformation of $\cos^2\xi\,=\,1-\mu^2$ and then $(y,\mu)$ to $(R,\Theta)$ defined by
\begin{equation}
\mu\,=\,1-\frac{1}{4}R^2\cos^2\frac{\Theta}{2}\,, \qquad y\,=\,y_1+\frac{4+5b+b^2+\left(4+3b\right)\sqrt{1+b}}{b^2y_1}R^2\sin^2\frac{\Theta}{2}\,.
\end{equation}
In the limit of $R\,\rightarrow\,0$, the metric becomes
\begin{align}
ds_{10}^2\,\approx&\,y_1^{-1/2}\left(ds_{AdS_2}^2+d\phi_3^2\right) \notag \\
+&\,dR^2+R^2\left\{\frac{d\Theta^2}{4}+\cos^2\frac{\Theta}{2}\left(d\psi^2+\sin^2\psi{d}\phi_1^2+\cos^2\psi{d}\phi_2^2\right)\right. \notag \\
+&\left.\mathcal{C}^2\left(1+b\right)\left[dz-\frac{1+\cos\Theta}{2\mathcal{C}\sqrt{1+b}}\left(\sin^2\psi{d}\phi_1+\cos^2\psi{d}\phi_2\right)\right]^2\right\}\,.
\end{align}
This is $AdS_3\,\times\,S^1_{\phi_3}\,\times\,\mathbb{R}^6/\mathbb{Z}_l$ and there is an orbifold singularity at the location of the monopole and is smooth elsewhere.

\bigskip

\noindent {\bf Region III: Smeared D3-branes} The singularity at $y\rightarrow\infty$ in the warp factor of five-dimensional metric, \eqref{wsingu}, has been resolved in the uplifted metric, \eqref{upmet1}. On the other hand, there is a singularity at $\left(y\rightarrow\infty,\,\sin\xi\rightarrow{0}\right)$ and we consider this singularity. We introduce coordinates, $(\rho,\Xi)$, for a reparametrization of $(y,\xi)$,
\begin{equation}
y\,=\,\frac{1}{\rho^{1/2}}\,, \qquad \sin\xi\,=\,\rho^{1/2}\cos\Xi\,.
\end{equation}
As $y\,\rightarrow\,\infty$ and $\sin\xi\,\rightarrow\,0$, or, equivalently, $\rho\,\rightarrow\,0$ and $\cos\Xi\,\rightarrow\,0$, the uplifted metric becomes
\begin{align}
ds_{10}^2\,\approx\,-&\frac{b}{4}\rho^{1/2}\cos\Xi\Big[ds_{AdS_3}^2+\mathcal{C}^2dz^2\Big] \notag \\
+&\frac{1}{\rho^{1/2}}\cos\Xi\left[\frac{1}{4}d\rho^2+\rho^2\left(d\phi_3^2+\sin^2\Xi\,d\Xi^2\right)+\frac{1}{\cos^2\Xi}\left(d\psi^2+\sin^2\psi{D}\phi_1^2+\cos^2\psi{D}\phi_2^2\right)\right]\,.
\end{align}
The metric implies the smeared D3-brane sources. The D3-branes are 
\begin{itemize}
\item extended along the $AdS_3$ and $z$ directions;
\item localized at the origin of the $\mathbb{R}^2$ parametrized by $S_{\phi_3}^1$ and $\rho$;
\item smeared along the $\Xi$ and $\left(\psi,\,\phi_1,\,\phi_2\right)$ directions. 
\end{itemize}
The $\rho^{1/2}$ factor of the space where the D3-branes are extended and the $1/\rho^{1/2}$ factor of the space where the D3-branes are localized and smeared corresponds to the harmonic functions of $H^{-1/2}$ and $H^{1/2}$ of the black D3-branes, respectively. See $e.g.$, appendix A of \cite{Couzens:2021rlk} for a brief review of smeared branes.

Around $y\,=\,0$, the five-form flux is given by
\begin{equation}
*G_{(5)}\,=\,-\text{vol}_{S^3}\wedge{d}\left(\cos^2\xi\right)\wedge{d}\phi_3\,.
\end{equation}
It indicates the existence of D3-brane source at $y\,=\,0$. Furthermore, the integral of it is finite and is calculated in \eqref{fluxq} below. Schematically, the source is of the form, $d*G_{(5)}\sim\delta(y)\,dy\wedge\text{vol}_{S^3}\wedge{d}\left(\cos^2\xi\right)\wedge{d}\phi_3$.

Lastly, we briefly present the comparison of our geometry with the geometry of wrapped M5-branes in \cite{Bah:2021mzw, Bah:2021hei}. The overall geometries are given by 
\begin{align} \label{compar}
\text{Wrapped D3-branes}: \qquad &AdS_3\,\times\,S_{\phi_3}^1\,\times\,S_z^1\,\times\,S^3(\psi,D\phi_1,D\phi_2)\,\times\,[y,\xi]\,, \notag \\
\text{Wrapped M5-branes}: \qquad &AdS_5\,\times\,\,S^2\,\,\times\,\,S_z^1\,\times\,S_\phi^1(D\phi)\,\times\,[w,\mu]\,,
\end{align}
where we denote the gauged coordinates with $D$, $e.g.$, $D\phi$. For each metric, we presented the factors in the same order so that the corresponding factors are easily found. For instance, in the geometry of M5-branes, there is one gauged coordinate, $D\phi$, in the circle of $S_\phi^1$. Therefore, there is a 3d base of $(w,\mu;\phi)$ fibered over $z$.  On the other hand, in the geometry of D3-branes, there are two gauged coordinates, $D\phi_1$ and $D\phi_3$, in the three-sphere of $S^3$. Therefore, there is the 5d base of $(y,\xi;\psi,\phi_1,\phi_2)$ fibered over $z$.

\subsection{Flux quantization}

We consider the flux quantization condition for the five-form flux. The integral of the five-form flux over any five-cycle in the internal space is an integer, see, $e.g.$, \cite{Couzens:2017way},
\begin{equation}
\frac{1}{\left(2\pi{l}_s\right)^4}\int_{M_5}F_{(5)}\,\in\,\mathbb{Z}\,,
\end{equation}
where $l_s$ is the string length.

First, we consider the $\left(*G_{(5)}\right)_{\xi\psi\phi_1\phi_2\phi_3}$ component of the five-form flux and we obtain{\footnote{A similar calculation was performed in appendix B of \cite{Arean:2008az}.}}
\begin{align} \label{fluxq}
\frac{1}{\left(2\pi{l}_s\right)^4}\int\left(*G_{(5)}\right)_{\xi\psi\phi_1\phi_2\phi_3}\,=&\,\frac{1}{\left(2\pi{l}_s\right)^4}\int\left(-\frac{2}{\Delta^2}\left(y^{2/3}+y^{-2/3}\Delta\right)\cos^3\xi\sin\xi\right)d\xi\wedge\text{vol}_{S^3}\wedge{d}\phi_3 \notag \\
=&\,-\frac{1}{4\pi{l}_s^4}\,\equiv\,-N\,,
\end{align}
where $N\,\in\,\mathbb{N}$ is the number of D3-branes wrapping the two-dimensional manifold, $\Sigma$. This integration contour corresponds to the interval, $\mathsf{Q}_1\mathsf{Q}_2$ in Figure 2.

Second, we consider the $\left(*G_{(5)}\right)_{\xi\phi_1\phi_3yz}$ and $\left(*G_{(5)}\right)_{\xi\phi_2\phi_3yz}$ components of the five-form flux and we obtain
\begin{align}
\frac{1}{\left(2\pi{l}_s\right)^4}\int\left(*G_{(5)}\right)_{\xi\phi_1\phi_3yz}\,=&\,\frac{1}{\left(2\pi{l}_s\right)^4}\int\left(-\frac{2\mathcal{C}}{y^3}\cos\xi\sin\xi\sin^2\psi\right)d\xi\wedge{d}\phi_1\wedge{d}\phi_3\wedge{d}y\wedge{d}z \notag \\
=&\,-\frac{1}{4\pi{l}_s^4}\frac{\mathcal{C}}{y_1^2}\sin^2\psi\,,
\end{align}
\begin{align}
\frac{1}{\left(2\pi{l}_s\right)^4}\int\left(*G_{(5)}\right)_{\xi\phi_2\phi_3yz}\,=&\,\frac{1}{\left(2\pi{l}_s\right)^4}\int\left(-\frac{2\mathcal{C}}{y^3}\cos\xi\sin\xi\cos^2\psi\right)d\xi\wedge{d}\phi_2\wedge{d}\phi_3\wedge{d}y\wedge{d}z \notag \\
=&\,-\frac{1}{4\pi{l}_s^4}\frac{\mathcal{C}}{y_1^2}\cos^2\psi\,.
\end{align}
In order to have identical results from these components, we fix $\psi\,=\,\frac{\pi}{4}$. Plugging $y\,=\,y_1$ from \eqref{regrange} and $l_s$ from \eqref{fluxq}, we obtain
\begin{align} \label{fluxk1}
\frac{1}{\left(2\pi{l}_s\right)^4}\int\left(*G_{(5)}\right)_{\xi\phi_1\phi_3yz}\,=\,\frac{1}{\left(2\pi{l}_s\right)^4}\int\left(*G_{(5)}\right)_{\xi\phi_2\phi_3yz}\,=\,-\frac{1}{4}\frac{N}{l}\frac{b}{1+b+\sqrt{1+b}}\,\equiv\,K\,,
\end{align}
where $K\,\in\,\mathbb{N}$ is another positive integer as $-1<b<0$.

From \eqref{clb}, \eqref{fluxq} and \eqref{fluxk1}, the parameters, $b$ and $\mathcal{C}$, are expressed in terms of the integers, $l$, $N$, and $K$, by
\begin{equation} \label{lnk1}
\sqrt{1+b}\,=\,\frac{N}{N+4Kl}\,, \qquad \mathcal{C}\,=\,\frac{N+4Kl}{Nl}\,.
\end{equation}

\subsection{Holographic central charge}

Now we calculate the holographic central charge of dual 2d superconformal field theories. For the metric of the form,
\begin{equation}
ds_{10}^2\,=\,e^{2\mathcal{A}}ds_{AdS_3}^2+ds^2_{M_7}\,,
\end{equation}
the formula for central charge is obtained in \cite{Couzens:2017way} from \cite{Brown:1986nw, Henningson:1998gx},
\begin{equation}
c\,=\,\frac{3}{2G_N^{(10)}}\int_{M_7}e^\mathcal{A}\text{vol}_{M_7}\,,
\end{equation}
where the ten-dimensional gravitational constant is $G_N^{(10)}\,=\,2^3\pi^6l_s^8$. Employing the formula, we obtain
\begin{equation} \label{cformula}
c\,=\,-\frac{3}{2}\frac{1}{2^3\pi^6l_s^8}\frac{b\,\mathcal{C}}{4}\pi^3\int_{y_{\text{min}}}^{y_{\text{max}}}\int_0^{2\pi}\frac{y}{\left(1-y^2\right)^2}dydz\,=\,-\frac{3b\,\mathcal{C}}{2^6\pi^2l_s^8}\left[\frac{1}{\left(1-y^2\right)}\right]_{y_{\text{min}}}^{y_{\text{max}}}\,,
\end{equation}
where $y_{\text{min}}\,=\,y_1$ and $y_{\text{max}}\,=\,\infty$ for the solutions in \eqref{regrange} and $\text{vol}_{S^5}=\pi^3$. We find the holographic central charge,
\begin{equation}
c\,=\,-\frac{3b\,\mathcal{C}}{2^6\pi^2l_s^8\left(y_1^2-1\right)}\,,
\end{equation}
where the minus sign is due to $-1<b<0$ and $1<y_1<\infty$. Employing the expressions for $l$ and $b$ from \eqref{lnk1}, we finally find
\begin{equation}
c\,=\,\frac{12lNK^2}{N+4Kl}\,.
\end{equation}
Even though the uplifted solutions have singularities, we obtain a well-defined finite result for central charge. 

\vspace{1.5cm}

\section{M5-branes wrapped on a product of topological disc and Riemann surface}

\subsection{Uplift formula}

Five-dimensional $SU(2)\times{U}(1)$-gauged $\mathcal{N}=4$ supergravity, \cite{Romans:1985ps}, is a consistent truncation of eleven-dimensional supergravity, \cite{Cremmer:1978km}, on $\Sigma_{\mathfrak{g}}\times{S}^4$, where $\Sigma_\mathfrak{g}$ is a Riemann surface of genus, $\mathfrak{g}$. In this section, we review the uplift formula, \cite{Gauntlett:2007sm}, in the conventions of \cite{Gauntlett:2007sm}.

The field content of five-dimensional $SU(2)\times{U}(1)$-gauged $\mathcal{N}=4$ supergravity, \cite{Romans:1985ps}, is the metric, an $SU(2)$ gauge field, $A^I$, $I\,=\,1,\,2,\,3$, a $U(1)$ gauge field, $A$, a complex two-form field, $B$, and a scalar field, $X\,=\,e^{-\frac{\phi}{\sqrt{6}}}$. The field strengths are defined to be
\begin{equation}
F\,=\,dA\,, \qquad F^I\,=\,dA^I-\frac{1}{\sqrt{2}}m\epsilon_{IJK}A^J\wedge{A}^K\,, \qquad C\,=\,dB+imA\wedge{B}\,,
\end{equation}
where we would set $m\,=\,1/2$.

Now we present the uplift formula. The uplift formula in \cite{Gauntlett:2007sm} is written in terms of functions of the scalar field, $X$, and the coordinate, $\xi$. We explicitly express the uplift formula in terms of $X$ and $\xi$. The fields in eleven-dimensional supergravity are the metric and the four-form flux. The uplift formula for the metric is
\begin{equation}
ds_{11}^2\,=\,\frac{\Omega^{1/3}}{2}ds_5^2+X\Omega^{1/3}\left(d\xi^2+ds_{\Sigma_\mathfrak{g}}^2\right)+\frac{2X}{\Omega^{2/3}}\sin^2\xi\left(d\psi+V+\frac{1}{2}A\right)^2+\frac{1}{X^2\Omega^{2/3}}\cos^2\xi{D}\mu^ID\mu^I\,,
\end{equation}
where we define
\begin{equation}
\Omega\,=\,2\cos^2\xi+\frac{1}{X^2}\sin^2\xi\,.
\end{equation}
The one-form, $V$, is defined by $dV\,=\,-\text{vol}_{\Sigma_{\mathfrak{g}}}$. The coordinates on a two-sphere are parametrized by $\mu^I$, $I\,=\,1,\,2,\,3$, with
\begin{equation}
D\mu^I\,=\,d\mu^I+\sqrt{2}m\epsilon_{IJK}A^K\mu^J\,.
\end{equation}
The ranges of the internal coordinates are
\begin{equation}
0\le\xi\le\frac{\pi}{2}\,, \qquad 0\le\psi\le2\pi\,.
\end{equation}
The uplift formula for the four-form flux is
\begin{align}
G_{(4)}\,=\,&-\frac{1}{8m^2}\epsilon_{IJK}\mu^ID\mu^JD\mu^K\wedge\left[-\frac{6X^2\cos^3\xi\sin^2\xi}{\left(2X^3\cos^2\xi+\sin^2\xi\right)^2}dX\wedge\left(d\psi+V+\frac{1}{2}A\right)\right. \notag \\
& \,\,\,\,\,\,\,\,\,\,\,\,\,\,\,\,\,\,\,\,\,\,\,\,\,\,\,\,\,\,\,\,\,\,\,\,\,\,\,\,\,\,\,\,\,\,\,\,\,\,\,\,\,\,\,\,\,\,\,\,\,\,\,\,\,\,\,\,\,\, +\frac{X^3\left(3+\cos(2\xi)\right)\cos^2\xi\sin\xi-\sin^5\xi}{\left(2X^3\cos^2\xi+\sin^2\xi\right)^2}d\xi\wedge\left(d\psi+V+\frac{1}{2}A\right) \notag \\
& \,\,\,\,\,\,\,\,\,\,\,\,\,\,\,\,\,\,\,\,\,\,\,\,\,\,\,\,\,\,\,\,\,\,\,\,\,\,\,\,\,\,\,\,\,\,\,\,\,\,\,\,\,\,\,\,\,\,\,\,\,\,\,\,\,\,\,\,\,\, \left.+2m\left(\cos\xi\,\text{vol}_{\Sigma_{\mathfrak{g}}}+\sin\xi\left(d\psi+V+\frac{1}{2}A\right)\wedge{d}\xi\right)\right] \notag \\
&+F^I\wedge\frac{1}{2\sqrt{2}m}\left[\frac{\cos\xi\sin^2\xi}{1+\left(2X^3-1\right)\cos^2\xi}D\mu^I\wedge\left(d\psi+V+\frac{1}{2}A\right)\right. \notag \\
& \,\,\,\,\,\,\,\,\,\,\,\,\,\,\,\,\,\,\,\,\,\,\,\,\,\,\,\,\,\,\,\,\,\,\,\,\,\,\,\,\, \left.-2m\mu^I\left(\cos\xi\,\text{vol}_{\Sigma_{\mathfrak{g}}}+\sin\xi\left(d\psi+V+\frac{1}{2}A\right)\wedge{d}\xi\right)\right] \notag \\
&+*_5F^I\wedge\left(-\frac{1}{4\sqrt{2}mX^2}\right)\left(\mu^I\sin\xi{d}\xi+\cos\xi{D}\mu^I\right) \notag \\
&+F\wedge\frac{1}{8m^2}\frac{X^3\cos^3\xi}{1+\left(2X^3-1\right)\cos^2\xi}\epsilon_{IJK}\mu^ID\mu^J\wedge{D}\mu^K \notag \\
&+C\wedge\frac{1}{4\sqrt{2}m}\sin\xi\left(e^1-ie^2\right)+c.c. \notag \\
&+B\wedge\frac{1}{2\sqrt{2}}\left(e^1-ie^2\right)\wedge\left(\cos\xi{d}\xi+i\sin\xi\left(d\psi+V+\frac{1}{2}A\right)\right)+c.c.\,,
\end{align}
where $\text{vol}_{\Sigma_{\mathfrak{g}}}\,=\,e^1\wedge\,e^2$ and $*_5$ denote the Hodge dual with respect to the five-dimensional metric, $ds_5^2$. The complex conjugates are denoted by $c.c$.{\footnote{We renamed a number of fields, $B^{\text{there}}\,\rightarrow\,A$, $C^{\text{there}}\,\rightarrow\,B$, and their field strengths, $G^{\text{there}}\,\rightarrow\,F$, $F^{\text{there}}\,\rightarrow\,C$, in \cite{Gauntlett:2007sm}. We also renamed some coordinates by $\theta^{\text{there}}\,\rightarrow\,\xi$, $x_3^{\text{there}}\,\rightarrow\,\psi$. The vielbeins, $e^1$ and $e^2$, are different from the ones in \cite{Gauntlett:2007sm}.}}

\subsection{Uplift to eleven-dimensional supergravity}

We identify the fields of five-dimensional $SU(2)\times{U}(1)$-gauged $\mathcal{N}=4$ supergravity (LHS) to our fields defined in section 2 (RHS) by
\begin{equation}
\phi=\phi_1\,, \qquad A^3=2\sqrt{2}A^1\,, \qquad A^3=2\sqrt{2}A^2\,, \qquad A^1=0\,, \qquad A^2=0\,, \qquad A=0\,, \qquad B=0\,.
\end{equation}
For the normalizations, see, $e.g.$, footnote 3 of \cite{Boido:2021szx}. Then, we are able to employ the uplift formula to uplift our solutions to eleven-dimensional supergravity.

By employing the uplift formula, we obtain the uplifted metric,
\begin{align} \label{upmet2}
ds_{11}^2\,=\,\frac{by^{4/3}\Omega^{1/3}}{8\left(1-y^2\right)}\left[ds_{AdS_3}^2+\frac{4}{y^2\left(1-y^2\right)h}dy^2+\frac{\mathcal{C}^2h}{b}dz^2+\frac{8\left(1-y^2\right)}{by^2}\left(d\xi^2+ds_{\Sigma_{\mathfrak{g}}}\right) \right. \notag \\
\left.+\frac{16\left(1-y^2\right)}{by^2\Omega}\sin^2\xi\left(d\psi+V\right)^2+\frac{8\left(1-y^2\right)}{b\Omega}\cos^2\xi{D}\mu^ID\mu^I\right]\,,
\end{align}
with
\begin{equation}
\Omega\,=\,2\cos^2\xi+y^2\sin^2\xi\,.
\end{equation}
We find the four-form flux to be
\begin{align}
G_{(4)}\,=\,&-\frac{1}{8m^2}\epsilon_{IJK}\mu^ID\mu^JD\mu^K\wedge\left[\frac{4y\cos^3\xi\sin^2\xi}{\left(2\cos^2\xi+y^2\sin^2\xi\right)^2}dy\wedge\left(d\psi+V\right)\right. \notag \\
& \,\,\,\,\,\,\,\,\,\,\,\,\,\,\,\,\,\,\,\,\,\,\,\,\,\,\,\,\,\,\,\,\,\,\,\,\,\,\,\,\,\,\,\,\,\,\,\,\,\,\,\,\,\,\,\,\,\,\,\,\,\,\,\,\,\,\,\,\,\, +\frac{y^2\sin\xi\left(2\cos^2\xi+2\cos^4\xi-y^2\sin^4\right)}{\left(2\cos^2\xi+y^2\sin^2\xi\right)^2}d\xi\wedge\left(d\psi+V\right) \notag \\
& \,\,\,\,\,\,\,\,\,\,\,\,\,\,\,\,\,\,\,\,\,\,\,\,\,\,\,\,\,\,\,\,\,\,\,\,\,\,\,\,\,\,\,\,\,\,\,\,\,\,\,\,\,\,\,\,\,\,\,\,\,\,\,\,\,\,\,\,\,\, +2m\left(\cos\xi\,\text{vol}_{\Sigma_{\mathfrak{g}}}+\sin\xi\left(d\psi+V\right)\wedge{d}\xi\right)\Big] \notag \\
&+F^I\wedge\frac{1}{m}\left[\frac{y^2\cos\xi\sin^2\xi}{y^2+\left(2-y^2\right)\cos\xi}D\mu^I\wedge\left(d\psi+V\right)\right. \notag \\
& \,\,\,\,\,\,\,\,\,\,\,\,\,\,\,\,\,\,\,\,\,\,\,\,\,\,\,\,\,\, -2m\mu^I\left(\cos\xi\,\text{vol}_{\Sigma_{\mathfrak{g}}}+\sin\xi\left(d\psi+V\right)\wedge{d}\xi\right)\Big] \notag \\
&+*_5F^I\wedge\left(-\frac{y^{4/3}}{2m}\right)\left(\mu^I\sin\xi{d}\xi+\cos\xi{D}\mu^I\right)\,.
\end{align}

The metric is schematically of the form,
\begin{equation}
AdS_3\,\times\,\Sigma(y,z)\,\times\,\Sigma_{\mathfrak{g}}\,\times\,S^2\,\times\,S^1_\psi\,\times\,[\xi]\,,
\end{equation}
which is the geometry of M5-branes wrapped on a product of topological disc and Riemann surface, $\Sigma(y,z)\,\times\,\Sigma_{\mathfrak{g}}$, with the internal four-sphere deformed to be $S^2\,\times\,S^1_\psi\,\times\,[\xi]$ reflecting the isometry of $SU(2)\,\times\,U(1)$. 

The singularity at $y=\infty$ in the warp factor of five-dimensional metric, \eqref{wsingu}, has been resolved in the uplifted metric, \eqref{upmet2}. On the other hand, there is a singularity at $\left(y\rightarrow\infty,\,\sin\xi\rightarrow{0}\right)$ and we consider this singularity. We introduce coordinates, $(\rho,\Xi)$, for a reparametrization of $(y,\xi)$,
\begin{equation}
y\,=\,\frac{1}{\rho^{1/2}}\,, \qquad \sin\xi\,=\,\rho^{1/2}\cos\Xi\,.
\end{equation}
As $y\,\rightarrow\,\infty$ and $\sin\xi\,\rightarrow\,0$, or, equivalently, $\rho\,\rightarrow\,0$ and $\cos\Xi\,\rightarrow\,0$, the uplifted metric becomes
\begin{align}
ds_{11}^2\,\approx\,-&\frac{b}{8}\rho^{1/3}\cos^{2/3}\Xi\Big[ds_{AdS_3}^2+\mathcal{C}^2dz^2+ds_{\Sigma_{\mathfrak{g}}}^2\Big] \notag \\
+&\frac{1}{\rho^{2/3}}\cos^{2/3}\Xi\left[\frac{1}{8}d\rho^2+R^2\Big(2\left(d\psi+V\right)^2+\sin^2\Xi\,d\Xi^2\Big)+\frac{1}{\cos^2\Xi}D\mu^I{D}\mu^I\right]\,.
\end{align}
The metric implies the smeared M5-brane sources. The M5-branes are 
\begin{itemize}
\item extended along the $AdS_3$, $z$ and $\Sigma_{\mathfrak{g}}$ directions;
\item localized at the origin of the $\mathbb{R}^2$ parametrized by $S_{\psi}^1$ and $\rho$;
\item smeared along the $\Xi$ and $\mu^I$ directions. 
\end{itemize}
The $\rho^{1/3}$ factor of the space where the M5-branes are extended and the $1/\rho^{2/3}$ factor of the space where the M5-branes are localized and smeared corresponds to the harmonic functions of $H^{1/3}$ and $H^{-2/3}$ of the black M5-branes, respectively.

Lastly, we briefly present the comparison with the geometry of wrapped D3-branes in \eqref{compar} and wrapped M5-branes in \cite{Bah:2021mzw, Bah:2021hei}. The overall geometries are given by 
\begin{align} \label{compar2}
\text{Wrapped D3-branes on $\Sigma$}:  \, \qquad \qquad AdS_3\,\times\,&S_{\phi_3}^1\,\times\,S_z^1\,\times\,S^3(\psi,D\phi_1,D\phi_2)\,\times\,[y,\xi]\,, \notag \\
\text{Wrapped M5-branes on $\Sigma\times{\Sigma}_{\mathfrak{g}}$}: \,\,\,\, \left(AdS_3\,\times\,\Sigma_{\mathfrak{g}}\right)\times\,&S^1_\psi\,\times\,S_z^1\,\times\,S^2\,\times\,[y,\xi]\,, \notag \\
\text{Wrapped M5-branes on $\Sigma$}: \qquad \qquad AdS_5\,\times\,\,&S^2\,\,\times\,\,S_z^1\,\times\,S_\phi^1(D\phi)\,\times\,[w,\mu]\,.
\end{align}
For each metric, we presented the factors in the same order so that the corresponding factors are easily found.

\subsection{Flux quantization}

We consider the flux quantization condition for the four-form flux. The integral of the four-form flux over any four-cycle in the internal space is an integer, see, $e.g.$, \cite{Boido:2021szx},
\begin{equation}
\frac{1}{\left(2\pi{l}_p\right)^3}\int_{M_4}G_{(4)}\,\in\,\mathbb{Z}\,,
\end{equation}
where $l_p$ is the Planck length.

First, we consider the four-form flux through $S^4\sim{S}^2\,\times\,S^1_\psi\,\times\,[\xi]$ and we obtain
\begin{equation} \label{fluxq2}
\frac{1}{\left(2\pi{l}_p\right)^3}\int_{S^4}G_{(4)}\,=\,\frac{1}{\pi{l}_p^3}\,\equiv\,N\,,
\end{equation}
where $N\,\in\,\mathbb{N}$ is the number of M5-branes wrapping a product of topological disc and Riemann surface, $\Sigma(y,z)\,\times\,\Sigma_{\mathfrak{g}}$.

Second, we consider the $\left(G_{(4)}\right)_{yz\psi\xi}$ component of the four-form flux and we obtain
\begin{align}
\frac{1}{\left(2\pi{l}_p\right)^3}\int\left(G_{(4)}\right)_{yz\psi\xi}\,=&\,\frac{1}{\left(2\pi{l}_p\right)^3}\int\left(-\frac{4\mathcal{C}}{y^3}\right)\left(-2\right)\left(\mu^1+\mu^2\right)\sin\xi\,dy\wedge{d}z\wedge{d}\psi\wedge{d}\xi\, \notag \\
=&\,\frac{1}{\pi{l}_p^3}\frac{2\mathcal{C}}{y_1^2}\left(\mu^1+\mu^2\right)\,,
\end{align}
Plugging $y\,=\,y_1$ from \eqref{regrange} and $l_p$ from \eqref{fluxq2}, we obtain
\begin{align} \label{fluxk}
\frac{1}{\left(2\pi{l}_p\right)^3}\int\left(G_{(4)}\right)_{yz\psi\xi}\,=\,-\frac{N}{l}\frac{b}{1+b+\sqrt{1+b}}\,\equiv\,K\,,
\end{align}
where we set $\mu^1+\mu^2\,=\,1$ and $K\,\in\,\mathbb{N}$ is another positive integer as $-1<b<0$.

From \eqref{clb}, \eqref{fluxq2} and \eqref{fluxk}, the parameters, $b$ and $\mathcal{C}$, are expressed in terms of the integers, $l$, $N$, and $K$, by
\begin{equation} \label{lnk}
\sqrt{1+b}\,=\,\frac{N}{N+Kl}\,, \qquad \mathcal{C}\,=\,\frac{N+Kl}{Nl}\,.
\end{equation}

\subsection{Holographic central charge}

Now we calculate the holographic central charge of dual 2d superconformal field theories. For the metric of the form,
\begin{equation}
ds_{11}^2\,=\,e^{2\mathcal{A}}\left(ds_{AdS_3}^2+ds^2_{M_8}\right)\,,
\end{equation}
the formula for central charge is obtained in \cite{Boido:2021szx} from \cite{Brown:1986nw, Henningson:1998gx},
\begin{equation}
c\,=\,\frac{1}{G_N^{(11)}}\int_{M_8}e^{9\mathcal{A}}\text{vol}_{M_8}\,,
\end{equation}
where the eleven-dimensional gravitational constant is $G_N^{(11)}\,=\,\frac{(2\pi)^8l_p^9}{16\pi}$. Employing the formula, we obtain
\begin{equation} \label{cformula}
c\,=\,\frac{16\pi}{(2\pi)^8l_p^9}\int_{y_{\text{min}}}^{y_{\text{max}}}\int_0^{2\pi}\frac{-b\,\mathcal{C}\,y}{8\left(1-y^2\right)^2}dydz\int\cos^2\xi\sin\xi\,d\xi\,d\psi\,\text{vol}_{\Sigma_{\mathfrak{g}}}\text{vol}_{S^2}\,=\,\frac{\mathfrak{g}-1}{12}\frac{N^3b\,\mathcal{C}}{1-y_1^2}\,,
\end{equation}
where $y_{\text{min}}\,=\,y_1$ and $y_{\text{max}}\,=\,\infty$ for the solutions in \eqref{regrange} and $\text{vol}_{S^2}=4\pi$ and $\text{vol}_{\Sigma_{\mathfrak{g}}}\,=\,4\pi\left(\mathfrak{g}-1\right)$. Employing the expressions for $l$ and $b$ from \eqref{lnk}, we finally find
\begin{equation}
c\,=\,\frac{lN^2K^2}{12\left(N+Kl\right)}\left(\mathfrak{g}-1\right)\,.
\end{equation}
Even though the uplifted solutions have singularities, we obtain a well-defined finite result for central charge.

\section{$\mathcal{N}\,=\,(4,4)$ supersymmetric $AdS_3$ solutions}

\subsection{Supersymmetry equations}

We consider the background,
\begin{equation}
ds^2\,=\,f(y)ds_{AdS_3}^2+g_1(y)dy^2+g_2(y)dz^2\,,
\end{equation}
with the gauge fields,
\begin{equation}
A^1\,=\,A^2\,=\,0\,, \qquad A^3\,=\,A_z(y)dz\,,
\end{equation}
and the scalar fields,
\begin{equation}
\phi_1\,=\,\varphi(y)\,, \qquad \phi_2\,=\,0\,.
\end{equation}

We solve the equation of motion for the gauge fields and obtain
\begin{equation}
A_z'\,=\,be^{\frac{4\varphi}{\sqrt{6}}}g_1^{1/2}g_2^{1/2}f^{-3/2}\,,
\end{equation}
where $b$ is a constant. As we proceed similar to the $\mathcal{N}\,=\,(2,2)$ case, we simply present the supersymmetry equations,
\begin{align}
0\,=\,&sf^{-1/2}\eta+g_1^{-1/2}\left[\frac{1}{2}\frac{f'}{f}+\frac{1}{2\sqrt{6}}\varphi'\right]\left(i\sigma^2\eta\right)+e^{-\frac{\varphi}{\sqrt{6}}}\left(\sigma^3\eta\right)\,, \notag \\
0\,=\,&g_2^{-1/2}\widehat{A}_z\left(\sigma^1\eta\right)-\frac{1}{2}f^{-3/2}e^{\frac{2\varphi}{\sqrt{6}}}\,b\,\eta-g_1^{-1/2}\left[\frac{1}{2}\frac{g_2'}{g_2}+\frac{1}{2\sqrt{6}}\varphi'\right]\left(i\sigma^2\eta\right)-e^{-\frac{\varphi}{\sqrt{6}}}\left(\sigma^3\eta\right)\,, \notag \\
0\,=\,&\frac{1}{3}\left(e^{-\frac{\varphi}{\sqrt{6}}}-e^{\frac{2\varphi}{\sqrt{6}}}\right)\left(\sigma^3\eta\right)+\frac{1}{2\sqrt{6}}g_1^{-1/2}\varphi'\left(i\sigma^2\eta\right)+\frac{1}{6}f^{-3/2}e^{\frac{2\varphi}{\sqrt{6}}}\,b\,\eta\,,
\end{align}
where $s\,=\,\pm1$.

\subsection{Supersymmetric solutions}

As we proceed similar to the $\mathcal{N}\,=\,(2,2)$ case, we simply present the final solution,
\begin{align}
f\,=&\,\frac{b}{2se^{-\frac{\varphi}{\sqrt{6}}}\left(e^{-\frac{\varphi}{\sqrt{6}}}-e^{\frac{2\varphi}{\sqrt{6}}}\right)}\,, \notag \\
g_1\,=&\,\frac{3b\left(\varphi'\right)^2}{8\left(e^{-\frac{\varphi}{\sqrt{6}}}-e^{\frac{2\varphi}{\sqrt{6}}}\right)^2\left(b-2se^{\frac{\varphi}{\sqrt{6}}}\left(e^{-\frac{\varphi}{\sqrt{6}}}-e^{\frac{2\varphi}{\sqrt{6}}}\right)\right)}\,, \notag \\
g_2\,=&\,\frac{\mathcal{C}^2\left(b-2se^{\frac{\varphi}{\sqrt{6}}}\left(e^{-\frac{\varphi}{\sqrt{6}}}-e^{\frac{2\varphi}{\sqrt{6}}}\right)\right)}{2se^{-\frac{\varphi}{\sqrt{6}}}\left(e^{-\frac{\varphi}{\sqrt{6}}}-e^{\frac{2\varphi}{\sqrt{6}}}\right)}\,, \notag \\
\widehat{A}_z\,=&\,\mathcal{C}e^{\frac{3\varphi}{\sqrt{6}}}\,,
\end{align}
where $\mathcal{C}$ is a constant. Therefore, we have determined all the functions in terms of the scalar field, $\varphi(y)$, and its derivative. The solution satisfies all the supersymmetry equations and the equations motion which we present in appendix A. We can determine the scalar field by fixing the ambiguity in reparametrization of $y$ due to the covariance of the supersymmetry equations,
\begin{equation}
\varphi(y)\,=\,\frac{2\sqrt{6}}{3}\log{y}\,,
\end{equation}
where $y\,>\,0$.

Finally, let us summarize the solution. The metric is given by
\begin{equation} \label{metmet2}
ds^2\,=\,\frac{by^{4/3}}{2s\left(1-y^2\right)}\left[ds_{AdS_3}^2+\frac{2}{sy^2\left(1-y^2\right)h(y)}dy^2+\frac{\mathcal{C}^2h(y)}{b}dz^2\right]\,,
\end{equation}
where we define
\begin{equation}
h(y)\,=\,b-2s\left(1-y^2\right)\,.
\end{equation}
The gauge field is given by
\begin{equation}
\widehat{A}_z\,=\,\mathcal{C}y^2\,.
\end{equation}

Now we consider the range of $y$ for regular solutions. We find regular solutions when we have
\begin{equation} \label{regrange2}
s\,=\,1\,, \qquad b>0\,, \qquad 1<y<y_1\,, \qquad y_1\,=\,\sqrt{1-\frac{b}{2}}\,.
\end{equation}
where $y_1$ is obtained from $h(y_1)\,=\,0$. We plot a representative solution with $b=-2$ and $\mathcal{C}\,=\,1$ in Figure 3.

\begin{figure}[t]
\begin{center}
\includegraphics[width=2.0in]{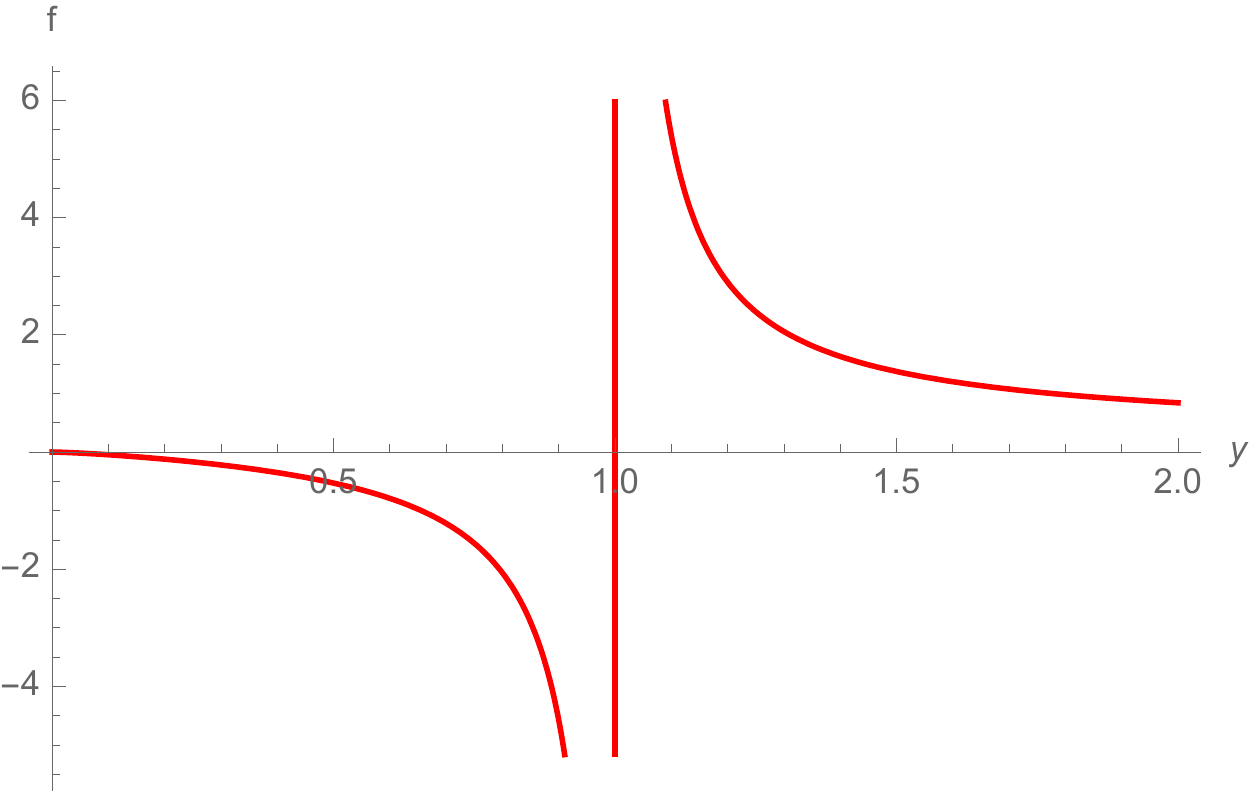} \qquad \includegraphics[width=2.0in]{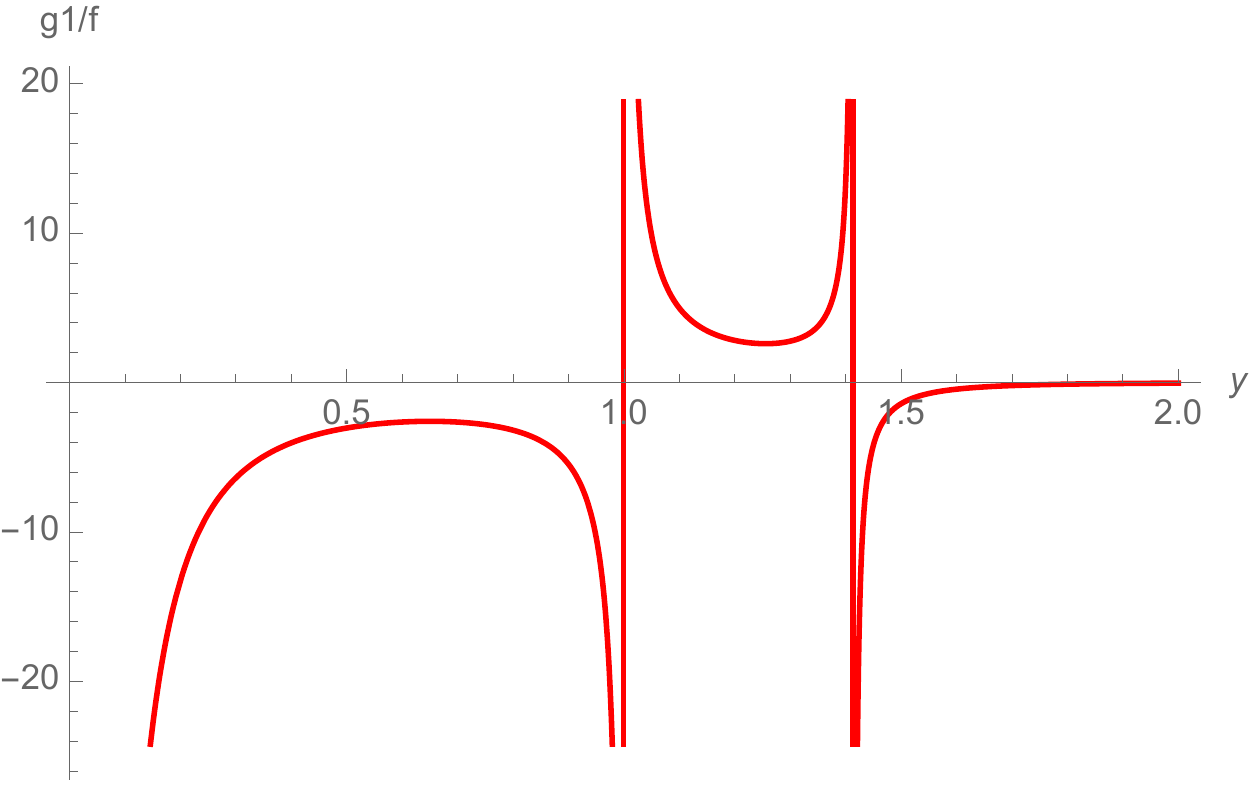} \qquad \includegraphics[width=2.0in]{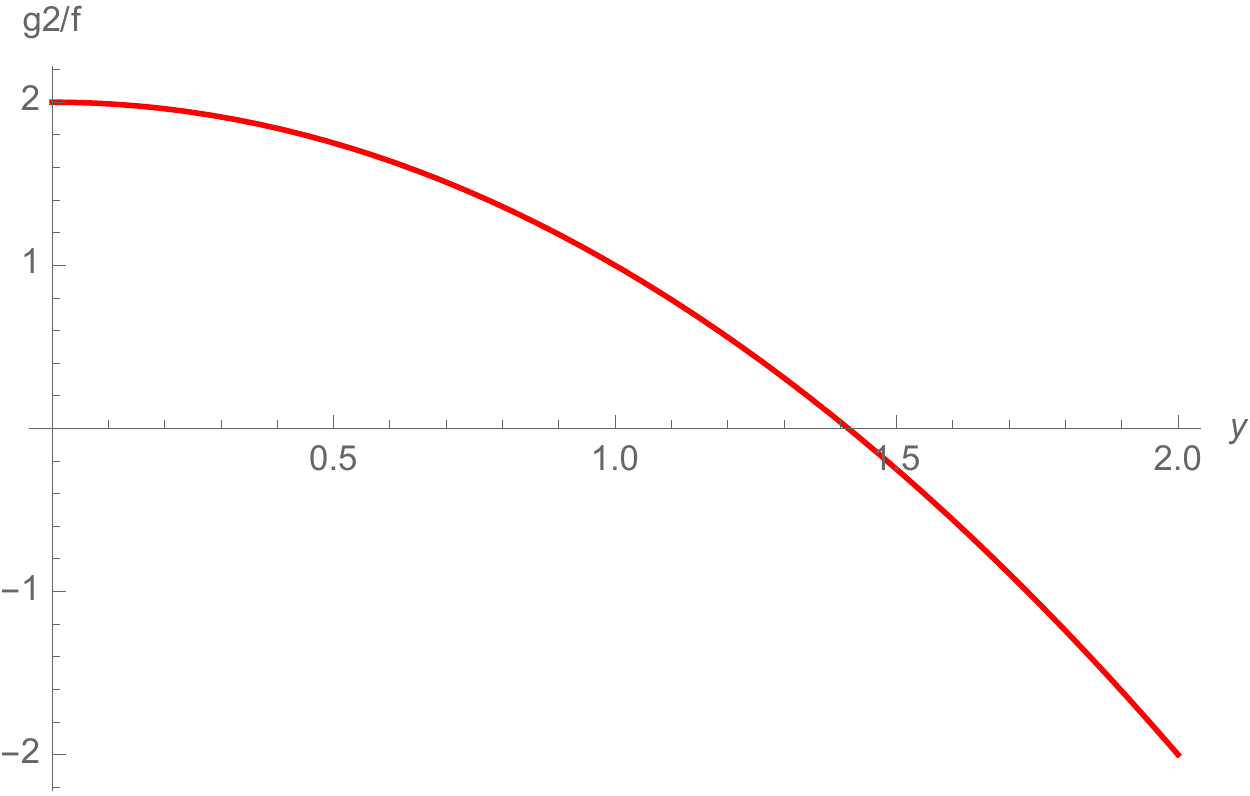}
\caption{{\it A representative solution with $b=-2$ and $\mathcal{C}\,=\,1$. The solution is regular in the range of $1\,<\,y\,<\,y_1=1.4142$.}}
\end{center}
\end{figure}

Approaching $y\,=\,y_1$, the metric becomes to be 
\begin{equation}
ds^2\,=\,\frac{by_1^{4/3}}{2s\left(1-y_1^2\right)}\left[ds_{AdS_3}^2+\frac{8\Big[d\rho^2+\frac{1}{4}\mathcal{C}^2\left(2-b\right)^2\rho^2dz^2\Big]}{-h'(y_1)y_1^2\left(1-y_1^2\right)}\right]\,,
\end{equation}
where we introduced a new parametrization of coordinate, $\rho^2\,=\,y_1-y$. Then, the $\rho$-$z$ surface is locally an $\mathbb{R}^2/\mathbb{Z}_l$ orbifold if we set
\begin{equation}
\mathcal{C}\,=\,\frac{1}{l\left(1-\frac{b}{2}\right)}\,,
\end{equation}
where $l\,=\,1,\,2,\,3,\ldots\,\,$.

Employing the Gauss-Bonnet theorem, we calculate the Euler characteristic of $\Sigma$, the $y$-$z$ surface, from \eqref{metmet2}. The boundary at $y\,=\,1$ is a geodesic and thus has vanishing geodesic curvature. The only contribution to the Euler characteristic is  
\begin{equation}
\chi\left(\Sigma\right)\,=\,\frac{1}{4\pi}\int_\Sigma{R}_\Sigma\text{vol}_\Sigma\,=\,\frac{2\pi}{4\pi}\frac{2\sqrt{2}\mathcal{C}y^3\left(y^2-1\right)}{\sqrt{by^2\left(1-y^2\right)}}\,=\,\mathcal{C}\left(1-\frac{b}{2}\right)\,=\,\frac{1}{l}\,,
\end{equation}
where $0\,<\,z\,<2\pi$. This result is natural for a disc in an $\mathbb{R}^2/\mathbb{Z}_l$ orbifold at $y\,=\,1$.

\subsection{Holographic central charge}

We uplift the solution to type IIB supergravity. The only non-trivial fields in type IIB supergravity are the metric and the RR five-form flux. The uplifted metric is already given in \eqref{upmet}. Furthermore, for the metric in \eqref{upmet}, the formula for holographic central charge is also given by \eqref{cformula}, which we repeat here for convenience,
\begin{equation}
c\,=\,-\frac{3b\,\mathcal{C}}{2^6\pi^2l_s^8}\left[\frac{1}{\left(1-y^2\right)}\right]_{y_{\text{min}}}^{y_{\text{max}}}\,,
\end{equation}
with $y_{\text{min}}\,=\,1$ and $y_{\text{max}}\,=\,y_1$ for the solutions in \eqref{regrange2}. However, the value of the central charge diverges at $y_{\text{min}}\,=\,1$. Unlike the $\mathcal{N}\,=\,(2,2)$ supersymmetric $AdS_3$ solutions, for the $\mathcal{N}\,=\,(2,2)$ case, we were not able to find any solution which do not have a bound at $y_{\text{min}}\,=\,1$.

Unlike the singularity of the warp factor, \eqref{wsingu}, which is resolved when uplifted, the singularity at $y=1$ is not resolved in the uplift. In order to avoid this singularity, we have to find solution which is well-defined in the range away from $y=1$. In the truncation we consider in this section, we were not able to find a solution of such range.

\section{Conclusions}

Employing the method applied to construct $AdS_5$ solutions from M5-branes recently by \cite{Bah:2021mzw, Bah:2021hei}, we constructed supersymmetric $AdS_3$ solutions from D3-branes and M5-branes wrapped on a topological disc with non-constant curvature. In five-dimensional $U(1)^3$-gauged $\mathcal{N}=2$ supergravity, we found $\mathcal{N}=(2,2)$ and $\mathcal{N}=(4,4)$ supersymmetric $AdS_3$ solutions. We uplifted the solutions to type IIB supergravity and obtained D3-branes wrapped on a topological disc. We also uplifted the solutions to eleven-dimensional supergravity and obtained M5-branes wrapped on a product of topological disc and Riemann surface. For the $\mathcal{N}=(2,2)$ solution, holographic central charges were finite and well-defined. On the other hand, we could not find $\mathcal{N}=(4,4)$ solutions with finite holographic central charge.

The first natural question would be to identify 2d $\mathcal{N}=(2,2)$ SCFTs which are dual to our solutions and compare the central charges with the gravitational results. Anomaly inflow method in \cite{Bah:2020jas} would be helpful to calculate the central charges. Furthermore, we would like to understand if there are indeed no $\mathcal{N}=(4,4)$ supersymmetric $AdS_3$ solutions with finite holographic central charge and understand the physics also in field theory.

In this work, we only constructed a class of $AdS_3$ fixed points from D3-branes and M5-branes on a topological disc. Holographic RG flows to the $AdS_3$ fixed points would enable us to understand more details of the solution. The uniformization problem was studied holographically in \cite{Anderson:2011cz, Bobev:2020jlb}.

In five-dimensional $U(1)^3$-gauged $\mathcal{N}=2$ supergravity, we would like to generalize the $AdS_3$ solutions to have both of the two scalar fields and all three $U(1)$ gauge fields to be non-trivial. This solution would be dual to 2d $\mathcal{N}=(0,2)$ SCFTs and generalize the result in \cite{Benini:2012cz, Benini:2013cda} where D3-branes are wrapped on a constant curvature Riemann surface. Extending to gauged $\mathcal{N}=2$ supergravity coupled to hypermultiplets would also be interesting.

The canonical form of supersymmetric $AdS_3$ solutions in type IIB supergravity only with the five-form flux is presented in \cite{Kim:2005ez}. We would like to embed our solutions in the canonical $AdS_3$ solution of type IIB supergravity.

Lastly, we demonstrated that the new way of realizing supersymmetry for solutions of gauged supergravity in \cite{Bah:2021mzw, Bah:2021hei} is a powerful tool. It would be very nice to see how we would apply the method to construct new supersymmetric solutions in the view toward deeper understanding of the AdS/CFT correspondence.

\bigskip
\leftline{\bf Acknowledgements}
\noindent We would like to thank Chris Couzens for interesting comments on the preprint and also an anonymous referee for instructive suggestions. This research was supported by the National Research Foundation of Korea under the grant NRF-2019R1I1A1A01060811.

\appendix
\section{The equations of motion}
\renewcommand{\theequation}{A.\arabic{equation}}
\setcounter{equation}{0} 

\subsection{Five-dimensional gauged $\mathcal{N}\,=\,2$ supergravity}

We present the equations of motion for gauged $\mathcal{N}\,=\,2$ supergravity coupled to two vector multiplets in five dimensions which we review in section 1. The equations of motion are given by
\begin{align}
R_{\mu\nu}&-\frac{1}{2}\partial_\mu\phi_1\partial_\nu\phi_1-\frac{1}{2}\partial_\mu\phi_2\partial_\nu\phi_2+\frac{4}{3}\left(e^{\frac{\phi_1}{\sqrt{6}}+\frac{\phi_2}{\sqrt{2}}}+e^{\frac{\phi_1}{\sqrt{6}}-\frac{\phi_2}{\sqrt{2}}}+e^{-\frac{2\phi_1}{\sqrt{6}}}\right)g_{\mu\nu} \notag \\
&-\frac{1}{2}e^{\frac{2\phi_1}{\sqrt{6}}+\frac{2\phi_2}{\sqrt{2}}}\left(F^1_{\mu\rho}F_\nu^{1\rho}-\frac{1}{6}g_{\mu\nu}F^1_{\rho\sigma}F^{1\rho\sigma}\right)-\frac{1}{2}e^{\frac{2\phi_1}{\sqrt{6}}-\frac{2\phi_2}{\sqrt{2}}}\left(F^2_{\mu\rho}F_\nu^{2\rho}-\frac{1}{6}g_{\mu\nu}F^2_{\rho\sigma}F^{2\rho\sigma}\right) \notag \\
&-\frac{1}{2}e^{-\frac{4\phi_1}{\sqrt{6}}}\left(F^3_{\mu\rho}F_\nu^{3\rho}-\frac{1}{6}g_{\mu\nu}F^3_{\rho\sigma}F^{3\rho\sigma}\right)\,=\,0,
\end{align}
\begin{align}
\frac{1}{\sqrt{-g}}\partial_\mu&\left(\sqrt{-g}g^{\mu\nu}\partial_\nu\phi_1\right)+\frac{4}{\sqrt{6}}\left(e^{\frac{\phi_1}{\sqrt{6}}+\frac{\phi_2}{\sqrt{2}}}+e^{\frac{\phi_1}{\sqrt{6}}-\frac{\phi_2}{\sqrt{2}}}-2e^{-\frac{2\phi_1}{\sqrt{6}}}\right) \notag \\
&-\frac{1}{2\sqrt{6}}\left(e^{\frac{2\phi_1}{\sqrt{6}}+\frac{2\phi_2}{\sqrt{2}}}F^1_{\mu\nu}F^{1\mu\nu}+e^{\frac{2\phi_1}{\sqrt{6}}-\frac{2\phi_2}{\sqrt{2}}}F^2_{\mu\nu}F^{2\mu\nu}-2e^{-\frac{4\phi_1}{\sqrt{6}}}F^3_{\mu\nu}F^{3\mu\nu}\right)\,=\,0\,, \notag \\
\frac{1}{\sqrt{-g}}\partial_\mu&\left(\sqrt{-g}g^{\mu\nu}\partial_\nu\phi_2\right)+\frac{4}{\sqrt{2}}\left(e^{\frac{\phi_1}{\sqrt{6}}+\frac{\phi_2}{\sqrt{2}}}-e^{\frac{\phi_1}{\sqrt{6}}-\frac{\phi_2}{\sqrt{2}}}\right) \notag \\
&-\frac{1}{2\sqrt{2}}\left(e^{\frac{2\phi_1}{\sqrt{6}}+\frac{2\phi_2}{\sqrt{2}}}F^1_{\mu\nu}F^{1\mu\nu}-e^{\frac{2\phi_1}{\sqrt{6}}-\frac{2\phi_2}{\sqrt{2}}}F^2_{\mu\nu}F^{2\mu\nu}\right)\,=\,0\,,
\end{align}
\begin{align}
D_\nu\left(e^{\frac{2\phi_1}{\sqrt{6}}+\frac{2\phi_2}{\sqrt{2}}}F^{1\nu\mu}\right)\,=\,0\,, \notag \\
D_\nu\left(e^{\frac{2\phi_1}{\sqrt{6}}-\frac{2\phi_2}{\sqrt{2}}}F^{2\nu\mu}\right)\,=\,0\,, \notag \\
D_\nu\left(e^{-\frac{4\phi_1}{\sqrt{6}}}F^{3\nu\mu}\right)\,=\,0\,.
\end{align}

\subsection{Type IIB supergravity}

The equations of motion of type IIB supergravity for the metric and self-dual five-form flux are
\begin{equation}
R_{MN}-\frac{1}{4}\left(|F_{(5)}|^2_{MN}-\frac{1}{2}|F_{(5)}|^2g_{MN}\right)\,=\,0\,,
\end{equation}
and
\begin{equation}
dF_{(5)}\,=\,0\,,
\end{equation}
where we define
\begin{equation}
|F_{(p)}|^2\,=\,\frac{1}{p!}F_{(p)M_1{\cdots}M_p}F_{(p)}\,^{M_1{\cdots}M_p}\,, \qquad |F_{(p)}|_{MN}^2\,=\,\frac{1}{(p-1)!}F_{(p)MM_1{\cdots}M_{p-1}}F_{(p)N}\,^{M_1{\cdots}M_{p-1}}\,.
\end{equation}

\section{Equivalence with the spindle}
\renewcommand{\theequation}{B.\arabic{equation}}
\setcounter{equation}{0} 

In this appendix, we show that the solution of the topological disc we obtain in \eqref{metmet} is, in fact, identical to the special case, \cite{Couzens:2021tnv}, of the multi-charge spindle solution in \cite{Hosseini:2021fge, Boido:2021szx}.

The multi-charge spindle solution in \cite{Boido:2021szx} is
\begin{align}
ds^2\,=&\,H(x)^{1/3}\left[ds^2_{AdS_3}+\frac{1}{4P(x)}dx^2+\frac{P(x)}{H(x)}d\phi^2\right]\,, \notag \\
A^I\,=&\,\frac{x-\alpha}{x+3K_I}d\phi\,, \notag \\
X^I\,=&\,\frac{H(x)^{1/3}}{x+3K_I}\,,
\end{align}
where the functions are given by
\begin{align}
H(x)\,=&\,\left(x+3K_1\right)\left(x+3K_2\right)\left(x+3K_3\right)\,, \notag \\
P(x)\,=&\,H(x)-(x-\alpha)^2\,,
\end{align}
and $K_1$, $K_2$, $K_3$, and $\alpha$ are constants. In \cite{Couzens:2021tnv} a special case was considered with
\begin{equation} \label{specials}
K_1\,=\,K_2\,=\,K\,, \qquad K_3\,=\,-\frac{1}{3}\alpha\,,
\end{equation}
where a constant, $K$, is introduced.

The topological disc solution we obtained in \eqref{metmet} is 
\begin{align}
ds^2\,=&\,\frac{by^{4/3}}{4\left(1-y^2\right)}\left[ds_{AdS_3}^2+\frac{4}{y^2\left(1-y^2\right)h(y)}dr^2+\frac{\mathcal{C}^2h(y)}{b}dz^2\right]\,, \notag \\
\widehat{A}_z\,=&\,\mathcal{C}\left(2y^{-2}-1\right)\,, \notag \\
\varphi\,=&\,\frac{2\sqrt{6}}{3}\log{y}\,,
\end{align}
where we introduced
\begin{equation}
h(y)\,=\,b-4y^{-4}\left(1-y^2\right)\,,
\end{equation}
and $b$ and $\mathcal{C}$ are constants. 

Now we perform a change of coordinates from $(x,\phi)$ to $(w,z)$ by
\begin{equation}
a^2w^2\,=\,x+3K\,, \qquad z\,=\,\phi\,,
\end{equation}
with a choice of the constants,
\begin{align}
b\,=\,-4a^2\,, & \qquad \mathcal{C}\,=\,1\,, \notag \\
3K\,+\,&\alpha\,=\,a^2\,,
\end{align}
where $a$ is a free parameter. Then we perform another change of the coordinate from $w$ to $y$ by{\footnote{We would like to thank Chris Couzens for an instructive comment on this.}}
\begin{equation}
y\,=\,w(w^2-1)^{-1/2}\,.
\end{equation}
Then both of the multi-charge spindle solution in the special case of \eqref{specials} and the topological disc solution we obtained here reduce to an identical solution,
\begin{align}
ds^2\,=&\,\frac{y^{4/3}}{y^2-1}\left[ds^2_{AdS_3}+\frac{dy^2}{y^2\left(y^2-1\right)\left(a^2-y^{-4}\left(y^2-1\right)\right)}+\frac{1}{a^2}\left(a^2-y^{-4}\left(y^2-1\right)\right)dz^2\right]\,, \notag \\
A^1\,=&\,A^2\,=\,2y^{-2}-1\,, \qquad A^3\,=\,0\,, \notag \\
X^1\,=&\,X^2\,=\,y^{-2/3}\,, \qquad X^3\,=\,y^{4/3}\,.
\end{align}



\end{document}